\documentclass[a4paper,11pt,oneside]{book}
\usepackage[T1]{fontenc}
\usepackage[utf8]{inputenc}
\usepackage{lmodern}
\usepackage[font=small,labelfont=bf]{caption}
\usepackage{amsmath}
\usepackage{sgame, tikz} 
\usepackage{algorithm2e}
\usepackage{hyperref}
\usepackage{graphicx}
\usepackage[english]{babel}

\title{\huge \textbf{Multi-issue Bargaining with \\ Deep Reinforcement Learning}  }
\author{Ho-Chun Herbert Chang}

\begin{document}

\frontmatter
\maketitle

\section*{Abstract}
Negotiation is a process where agents aim to work through disputes and maximize their surplus. As the use of deep reinforcement learning in bargaining games is unexplored, this paper evaluates its ability to \textbf{exploit}, \textbf{adapt}, and \textbf{cooperate} to produce fair outcomes, in comparison to classical game theoretic results.

Two actor-critic networks were trained for the bidding and acceptance strategy, against time-based agents, behavior-based agents, and through self-play. Gameplay against these agents reveals three key findings. 1) Neural agents learn to \textbf{exploit} time-based agents, achieving clear transitions in decision preference values. The Cauchy distribution emerges as suitable for sampling offers, due to its peaky center and heavy tails. The kurtosis and variance sensitivity of the probability distributions used for continuous control produce trade-offs in exploration and exploitation. 2) Neural agents demonstrate \textbf{adaptive behavior} against different combinations of concession, discount factors, and behavior-based strategies. 3) Most importantly, neural agents learn to \textbf{cooperate} with other behavior-based agents, in certain cases utilizing non-credible threats to force fairer results. This bears similarities with reputation-based strategies in the evolutionary dynamics, and departs from equilibria in classical game theory.

\tableofcontents

\mainmatter







\chapter{Introduction}

Negotiation is the process where parties interact to settle issues, discover surplus, and create contracts. Because negotiation is so essential to society, it has been widely studied by different fields, in economics~\cite{osborne1994course, raiffa1982art},  artificial intelligence~\cite{jennings2001automated, kraus1997negotiation, gerding2000scientific}, business~\cite{walsh1999modeling,lewicki2016essentials,ehlich2011discourse,huang2010agent}, communication~\cite{arvanitis2011negotiation,maaravi2011negotiation}, and behavioral psychology~\cite{rubin2013social,de2007psychology}. 

However, negotiation is typically very costly. Automated negotiation is a field that, for the past 20 years, has promised to reduce the costs of human negotiation, avoid social confrontation, and augmented the abilities of human negotiators. 
It has found great success in e-commerce and supply chain management~\cite{fang2008opponent}. 
Recent success in deep reinforcement learning (DRL) in games, such as Chess~\cite{silver2016mastering}, Go~\cite{silver2017mastering}, Poker~\cite{brown2018superhuman} and Atari games~\cite{mnih2013playing} have inspired DRL's application to complex human tasks, including negotiation~\cite{lewis2017deal}. However, the direct application of DRL on negotiation, formally known as bargaining games within game theory, is unexplored. The central goal of this dissertation is to understand the behavioral dynamics of negotiation bots trained with DRL, and assess its ability to \textbf{exploit}, \textbf{adapt},  and \textbf{cooperate}.

Beyond reducing costs, understanding the behavior of DRL agents is paramount because, with the deployment of any technology, there comes potential harm. When DRL was applied to natural language generation for negotiation, agents were shown to misrepresent their intentions within negotiation dialogues~\cite{lewis2017deal}. As natural language generation techniques mature, it is easy to imagine a future where complex transactions, contracts, and negotiations are facilitated through bots. 
Already, more than 80\% of WeChat transactions are mediated through chatbots and conversational marketing--- the use of customized chatbots in marketing---  has been forecasted as the most dominant interface between consumers and businesses~\cite{eeuwen2017mobile}. 
As it is well-established that suppliers hold more information that buyers~\cite{akerlof1978market}, DRL bots may exploit asymmetries when mediating markets, exposing consumers to representational harm.


To be explicit about our problem, Baarslag divides negotiation into three pillars--- the \textit{bidding strategy}, \textit{opponent modeling}, and \textit{acceptance strategy}. This dissertation evaluates the extent in which actor-critic models can learn end-to-end bidding and acceptance strategies through gameplay, solely through the opponent's behaviors alone. As mentioned prior, the three behavioral traits of interest are \textbf{exploitation} (ability to utilize an opponent's weaknesses), \textbf{adaptation} (ability to play against different opponents), and \textbf{cooperation} (ability to work with other agents for fairer and more efficient results).
Related sub-problems include 1) when sub-optimal bidding results occur, 2) what the variables that produce these limitations are, and 3) how neural agents learn fair strategies against more complicated agents. The emergence of fairness is often studied through the lens of game theory~\cite{von2007theory, verbeek2004game}; we adopt a similar approach for analysing our agent's behavior.

This project aims predominately to contribute to \textbf{automated negotiation}. We contribute by implementing DRL within the bargaining domain, through competition against time-based agents, tit-for-tat agents, and self-play. The first half of the results focuses on precision--- the agent's mechanics and exploitative abilities, whereas the second half on behavioral dynamics.
There are three key results for behavior. 
First, we demonstrate exploitative and adaptive behavior against time-based agents and the simple behavioral agent, showing clear switching behavior in both acceptance and bid strategy, up to 80\% and 90\% (Pareto) efficiency respectively. 
Second, the neural network cooperates with more complex behavior-based agents to produce fairer results. Furthermore, in a game where agents are allowed to accrue interest, it learns to cooperate during self-play. Third, and most interestingly, results demonstrate the emergence of fairness through the adoption of non-credible threats This is considered irrational in classic game theory. We draw a comparison to results in evolutionary game theory, where a similar divergence from assumptions of rationality.

Additionally, we contribute an analysis that features contemporary reinforcement learning research directions~\cite{arulkumaran2017brief}--- negotiation with deadlines is synonymous with the cliff-walking problem and construing offers aligns with control over continuous state space. Using derivative-based analysis based on marginal utility for optimal stopping, I demonstrate a negative relationship between time error and the second derivative. Secondly, successful continuous control on this domain requires careful control over exploration and exploitation--- the Cauchy distribution arises as a candidate due to its peaky center, heavy tails, and sensitivity to change in variance.


\chapter{Background and Terminology}

Due to the expensive nature of negotiation, attention to automating the process has gained considerable traction in the past twenty years~\cite{baarslag2012first}, since the development of Contract Net Protocol by Smith in the 1980s~\cite{smith1980contract}. It spurned the promise of finding better outcomes than human negotiators~\cite{bosse2005human, dzeng2005searching, jazayeriy2011review,oshrat2009facing,vahidov2014experimental,lin2010can}.

Baarslag~\cite{baarslag2016learning} identifies three key aspects of negotiation strategy--- \textit{bidding strategy}, \textit{opponent modeling}, \textit{acceptance strategy}. The bidding strategy asks what concessions to make in the process of counter-offering. The acceptance strategy asks when and which offers should be accepted. Opponent modeling focuses on understanding what the opponent wants, in order to make better decisions about bidding and acceptance.

	Negotiation is simultaneously collaborative and competitive. A good negotiation outcome is characterized by 1) win-win situations 2) avoiding no agreement, 3) and avoiding exploitation~\cite{baarslag2016learning}. While the first two activities are collaborative, the third is competitive. Negotiators are typically unwilling to share information in fear of being exploited~\cite{niemann2009assess,coehoorn2004learning,raiffa1982art}. Thus, one main challenge of negotiation is overcoming the information barrier, which makes bargaining a game of incomplete information. 

    Therefore, opponent modeling can be regarded as the most important in the field of negotiation
. Like poker, successful negotiation arises from understanding your opponent and generating profits off their behavioral heuristics and weaknesses. Especially given the diversity of negotiation agents, it is difficult to produce a singular agent that plays well against multiple opponent typologies. Therefore, a focus of the field is less on \textit{optimal play}, but \textit{exploitative play} using adaptive agents~\cite{baarslag2016learning}.
    
    
    Before reviewing the state-of-the-art, we layout the necessary terminology for negotiation, beginning with the mechanism,  measures of fairness and optimality, then discuss the common negotiation strategies used to benchmark negotiation models.
    

\section{Overview of Negotiation}
A \textit{negotiation setting} contains a \textit{protocol}, \textit{agents}, and \textit{scenario}. The protocol determines the rules of how agents interact with each other. The \textit{scenario} takes place in a \textit{negotiation domain} which determines an \textit{outcome space}, denoted as $\Omega$. A negotiation domain can have a single or multiple issues. \textit{Issues} refer to the resources under contention, such as the price of an object or level of service. Thus, an outcome $\omega \in \Omega$ can be described as a specific division of the issues. 
Agents have \textit{preference profiles}, which determines specific outcomes they prefer.

\subsection{Protocols}
We use single-issue bargaining as a preliminary illustration. Given a unit pie, two players $A$ and $B$ are asked to split it amongst themselves~\cite{fatima2013negotiation}. Suppose Agents $A$ and $B$ negotiate $N$ rounds to divide a unit pie, by alternately proposing outcomes called \textit{bids} or \textit{offers}, until a player accepts. We denote an offer $x = (x, y)$, such that $x +  y = 1$. 

This process of alternating offers is known as the Rubenstein's Bargaining Protocol. Games with one round, are known as an \textit{ultimatum game}~\cite{rubinstein1982perfect}. In ultimatum games, $A$ makes the first and only proposal. $B$ can only accept or reject it, which means $A$ has all the power. Similarly, if there are two rounds, then Player $B$ has the advantage. 
In a game of repeated offers, it is necessary to introduce some form of discounting factor--- otherwise, players would negotiate forever. The discount factor $\delta$ makes a portion of the pie go bad at every round.  Thus, it is in the best interest for players to finish the game as soon as possible.

The Rubenstein Bargaining Protocol is widely used because it accurately simulates many real-world scenarios~\cite{rubinstein1982perfect}. 
Multi-issue bargaining is more complex, as multiple issues are under contention and requires further protocol restrictions describe how each issue is resolved. Common ones are~\cite{kraus1997negotiation}:
\begin{enumerate}
    \item \textbf{Package-deal Procedure: } All issues addressed at once. 
    \item \textbf{Simultaneous Procedure: } All issues are solved independently. It is equivalent to $m$ single-issue problems.
    \item \textbf{Sequential Procedure: } Negotiates one issue at a time, with a predetermined sequence. Cannot negotiate prior or future issues.
\end{enumerate}

An alternative protocol is the \textit{monotonic concession protocol}~\cite{rosenschein1994rules}, where agents disclose information about how they value each issue, and their subsequent offers must have less utility than their prior ones. Other protocol considerations include~\cite{fatima2014principles}:
\begin{enumerate}
    \item \textbf{Time Constraints: } Beyond the discount factor $\delta$, there is often a deadline $T$. If negotiation does not end by $T$, players earn $0$ utility (known as the \textit{conflict deal}).
    \item \textbf{Divisibility:} Issues may be atomic and discrete, or divisible and continuous.
    \item \textbf{Lateral-ness:} Whether negotiation is between two parties (bilateral) or with multiple parties (multilateral).
    \item \textbf{Reserve Price $r$:} The minimum an agent is willing to accept.
\end{enumerate}

\subsection{The Scenario}
The utility is defined as the $\textit{cumulative utility}$,: a combination of sub-utility functions. Most commonly used is the linear additivity. With $\vec{x} $ the division for Player A (PA) and $\vec{y}$ for Player B (PB), the aggregate utility is of PA is:

\begin{equation}
U_1(x,t) = 
\begin{cases}
    \delta^{t-1} \vec{W}^T \vec{x}  = 
    \displaystyle\sum_{i=1}^m w_{i,a} x_i \delta^{t-1}
    & \text{if } t \leq T  \\
    0 & \text{otherwise (Conflict Deal)}
\end{cases}
\end{equation}
where $w_{i,a}$ is the value (weight) PA ascribes to issue $i$, $\delta$ the discount rate, and $x_i$ the division for issue $i$. This can be viewed as the discounted dot product of weights $\vec{W}$ and issue division $x$. 
 In many cases, however, utilities are not linear in combination--- for instance, in the auctions of multiple items, combinations of items yield greater rewards, to the effect of the sum being greater than the parts, due to synergistic effects. These are modeled with \textit{non-linear utility functions}~\cite{ito2008multi}.

The action space is defined by three possible actions: $A_i = \{ \text{Offer(x,y)}, \\ \text{reject},\text{accept} \}$. Offers are made after rejections, and should an agent choose to accept an offer the negotiation ends. Each issue is often normalized such that $x_i + y_i = 1$. For games with only one issue, the offer consists of the division of one pie. For multiple issues, offers are represented as vectors, subject to $\vec{x} + \vec{y} = 1$. For this dissertation, the outcome space is assumed to continuous, linear, and normalized.

\subsection{Outcome Spaces}
Each player has a preference ordering, called the \textit{preference profiles}, on all possible outcomes. An outcome $\omega ' $ is weakly preferred to $\omega$ if $u(\omega') \geq u(\omega)$, which is denoted $\omega' \geq \omega$. Similarly, $\omega ' $ is strictly preferred to $\omega$ (denoted $\omega' > \omega$) if $u(\omega') >  u(\omega)$. For linear additive utilities, the preference profile can be inferred directly by the weights. 

Now we present metrics used to evaluate our three criterion.
An outcome is called \textit{Pareto Optimal} if there exists no outcome $\omega'$ that a player would prefer without worsening their opponent's outcome. Formally:
$$
(\omega' >_A \omega \land \omega' \geq_B \omega) ~\lor ~
(\omega' >_B \omega \land \omega' \geq_A \omega) 
$$
The \textit{Pareto Frontier} describes all Pareto optimal solutions, which we denote as $\Omega_P$. When an offer is not Pareto Optimal, then through negotiation there is potential to reach an outcome without players conceding anything.

There are two other useful metrics. Let $\omega_P \in \Omega_P$ denote the set of outcomes that are Pareto optimal. The \textit{bid distribution} denotes the mean distance to the Pareto frontier, shown in Eq.~\ref{eq:bid-distrib}. A high bid distribution indicates bids are on average far away.
\begin{equation} \label{eq:bid-distrib}
    BD(\Omega) = \sum_{\omega \in \Omega} \frac{\min{d( \omega, \omega_P)} } {|\Omega |}
\end{equation}{}

Usually, simultaneous maximization of outcomes is not possible, as there is a region of disagreement between players. Another useful metric is the product of utilities ($U_A \cdot U_B$), known as the \textit{Nash Product}. A fair outcome is often characterized using the \textit{Nash solution}, the outcome that maximizes the product of utilities, shown in Eq.~\ref{eq:nash-solution}.
\begin{equation} \label{eq:nash-solution}
    \omega_{Nash} = \max_{\omega \in \Omega} U_A(\omega) \cdot U_B(\omega)
\end{equation}{}

\section{Strategies}
	In cases of perfect information, it is possible to determine the optimal bidding strategy~\cite{fatima2014principles}.
However, as previously mentioned, perfect information is unlikely in bargaining as agents are unwilling to give away their preferences in fear of exploitation. This motivates the development of negotiation tactics under imperfect information.
    These negotiation tactics can broadly be classified as \textit{time-dependent} or \textit{behavior-dependent} tactics, based on a \textit{decision-function} that maps state to a target utility.

\subsection{Baseline Strategies}
Two are commonly used. The \textit{Hardliner} always bids maximum utility for itself, which emulates the "take-it-or-leave-it" attitude. The \textit{Random walker} denotes agents that bid randomly, serving as a standard baseline.

\subsection{Time-dependent Strategies}
\begin{figure}[!htb]
    \centering
    \includegraphics[width = 0.8\linewidth]{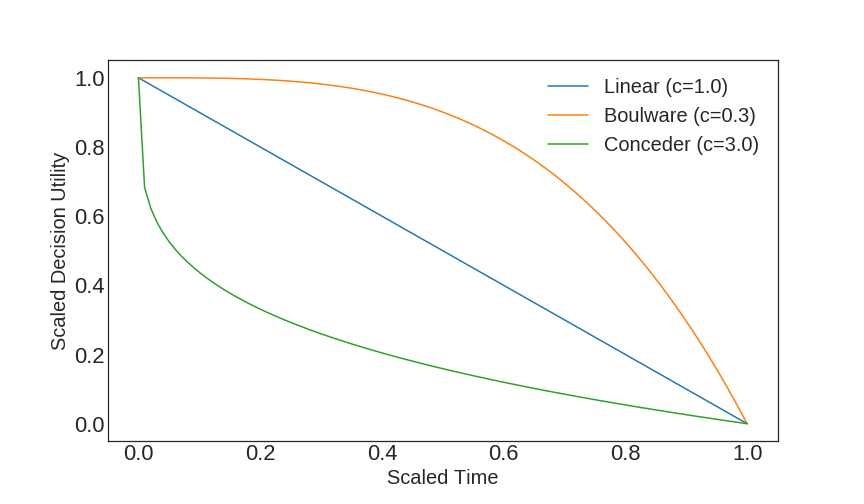}
    \caption{Decision utilities of time-based agents with different concession factors.}
    \label{fig:time-based-agents}
\end{figure}

Time-dependent Strategies denote functions that produce offers solely based on time. At every round, the agent calculates their \textit{decision utility} which determines whether they accept an offer or not. For time-dependent agents, this is:
\begin{equation}\label{eq:decision-util}
    u(t) = P_{min} + (P_{max} - P_{min}) \dot (1-F(t))
\end{equation}
$P_{max}$ and $ P_{min} \in [0,1]$, thus parametrizing the range of the offers. Frequently, $F(t)$ is parametrized as an exponential function:
\begin{equation} \label{eq:concession-factor}
    F(t) = k + (1-k) \cdot \big( \frac{t}{T} \big)^{1/c}
\end{equation}
where $c$ is the concession factor. $k$ is often set to 0 for simplicity. Fig.~\ref{fig:time-based-agents} shows the decision utilities of different agents. If $0 < c < 1$, then the agent concedes towards the end and is known as \textit{Boulware}. Otherwise, if $c \geq 1$, the agent concedes quickly and offers its reservation value, thus it is known as a \textit{Conceder}.  $c = 1$ means the agent's decision utility decreases linearly.

\subsection{Behavior-based Strategies} \label{sec:behavior-based}
\textit{Behavior-dependent} and \textit{imitative} bidding strategies observe the behavior of the opponent to make their own decisions on what to offer and what to accept. The most well-known is \textit{tit-for-tat}, which produces cooperation through reciprocity. It's three central mantras are 1) never defect first (play nice as long as the opponent plays nice), 2) retaliate if provoked and 3) can forgive after retaliation.

In negotiation, the \textit{relative tit-for-tat} (TFT) strategy reciprocates by offering concessions proportional to their opponent's concessions from $\delta$ rounds prior:
\begin{equation}\label{eq:t4t-decision-funct}
    x^{t_{n+1}}_{a \rightarrow b} [j] = \min \Bigg( 
\max \Bigg( 
\frac{x^{t_{n-2\delta}}_{b \rightarrow a} [j]}{x^{t_{n-2\delta+2}}_{b \rightarrow a} [j]}
x^{t_{n-1}}_{a \rightarrow b} [j] , \min_j^a
\Bigg), \max^a_j
\Bigg)
\end{equation}
Here, $x^{t_{n+1}}_{a \rightarrow b} [j] $ is the offer for issue $j$. This value is determined by the ratio of the opponent's prior concessions, which then scales the agent's own prior offer 
$x^{t_{n-1}}_{a \rightarrow b} [j] $. The min and max values ensure offer values are within range.

%
\section{State-of-the-Art in Negotiation}
Machine learning methods in the domain of negotiation can be broadly separated into the following types: Bayesian learning, non-linear regression, kernel density estimation, and artificial neural networks. These methods have been applied to mostly model an opponent's (acceptance and bidding) strategy, then derive an analytic response. This is because if an agent knows the opponent's bidding strategy, then the agent can compute its optimal strategy
~\cite{baarslag2016learning}.

For estimating the opponent acceptance strategy, techniques can be siloed into the estimation of individual variables. Zeng and Sycara provide a popular and intuitive Bayesian approach for estimating the reserve price, using historical data. The model generates a set of hypotheses on the opponent's reserve price, then attaches a likelihood using the history. The estimate is a weighted sum of the hypotheses based on their likelihoods~\cite{zeng1998bayesian}. This technique has been adapted to estimate the deadline for time-dependent tactics~\cite{sim2008blgan}.  In general, acceptance strategy estimation uses some form of Bayesian learning~\cite{sycara1997benefits,yu2013adaptive,sim2007adaptive,gwak2010bayesian,ren2002learning}, augmented with non-linear regression~\cite{agrawal2009learning,yu2013adaptive, sim2008blgan,haberland2012adaptive,hou2004predicting}, kernel density estimates~\cite{farag2010towards,oshrat2009facing,coehoorn2004learning}, polynomial interpolation~\cite{saha2005modeling}, genetic algorithms~\cite{matwin1991genetic,jazayeriy2011learning}, and more recently neural networks~\cite{fang2008opponent}.

In contrast, neural methods have been applied much more aggressively to the bidding strategy~\cite{baarslag2016learning}. In simpler cases where the general bidding formula is known, regression is sufficient as the problem reduces down to parameter estimation. If no formula is known, then neural networks are employed to approximate the opponent's bid strategy, typically using a large database of bid history. Oprea~\cite{oprea2002adaptive} uses a time-series approach on single-issue negotiations, taking in only the opponent's current bid. 
By 2008, early efforts for opponent move prediction using neural networks~\cite{carbonneau2008predicting}, who focused on predicting human bidding strategies. This was particularly relevant in e-commerce and supply chain management, as forecasting bids is useful in determining automated strategies~\cite{lee2009neural, carbonneau2011pairwise, moosmayer2013neural}. When the domain is general, researchers have found success using deep learning with multilayer perceptrons. Masvoula shows reliable predictions using single deep networks both with and without historical knowledge~\cite{masvoula2005design,masvoula2011predictive}.  Papaioannou and Rau et al. have shown the concession factor and weight of each issue can be predicted if the opponent is known to be time-dependent, using multilayer neural nets~\cite{papaioannou2008neural,rau2006learning} or single layer, radial basis function neural nets~\cite{papaioannou2011multi}.

Reinforcement learning approaches to negotiation began as early as the late 20th century, often denoted as adaptive learning~\cite{rapoport1998reinforcement}. Today DRL has more frequently been used with natural language processing~\cite{georgila2011reinforcement,cuayahuitl2015strategic}. 
Lewis et al. implement an end-to-end DRL negotiation dialogue generator~\cite{lewis2017deal}. They curated a set of human-human dialogues with Mechanical Turk, then trained on four gated recurrent units~\cite{cho2014properties}, a type of long-short term memory neural net~\cite{neubig2017neural}.  However, this study focuses on emulating human language, with less concern on optimality--- for instance, their DRL agent present 58.6\% and 69.1\% Pareto Optimality against simple autonomous agents and humans respectively, on a very limited, discrete action space (around 200 offers).

As illustrated with this brief survey, there is an immense number of agent designs for negotiation. The primary weakness in best performing models, such as Bayesian models in acceptance or bid prediction, is they require specific domain assumptions and architectures. Another weakness is for these negotiators to perform well in populations of different strategies, an additional opponent classifier is needed, which introduces further uncertainty. Additionally, opponents can use more complex behavioral strategies and mixed strategies--- pure strategies associated with a probability---that requires higher levels of adaptability to play against.

All of this motivates an adaptive agent with a fixed architecture that can perform well against different opponents. An end-to-end negotiation agent is desirable as the only required input is the offer, time step, and public knowledge, and can adapt online during gameplay.
Although deep learning often comes at the expense of explainability, a fixed architecture playing end-to-end means we do not need additional classifiers and assumptions about the opponent. The success of AlphaZero in chess is largely because it did not rely on hand-crafted heuristics and assumptions like other engines~\cite{silver2017mastering}; likewise, Libratus learned to exploit specific human opponent idiosyncrasies in poker~\cite{brown2018superhuman}. An end-to-end, adaptive neural agent is the analogous solution for negotiation. It is a convenient coincidence that the negotiation domain also aligns with the current interest in continuous control positions deep reinforcement learning.

\chapter{Methods}
Each bilateral negotiation scenario takes these as an input: \textit{two utility weights} $w_1$ and $w_2$, discount factors $\delta$, a deadline of $20$, reserve price of $0$, and two agents. These two agents then negotiate to an agreement or disagreement (conflict deal). For simplicity, our agent's utility weights are $(1,2,3)$ whereas the opponent's is $(3,2,1)$. Experiments consist of the neural agent playing against a second agent until reaching stopping criteria.

\section{Deep Reinforcement Learning}
Multi-agent Reinforcement Learning is formally the study of n-agent stochastic games~\cite{shoham2003multi}, described as a tuple $(N,S, \vec{A}, \vec{R}, T)$. $N$ is the number of agents. $S$ is the set of states and $\vec{A} = A_1,…,A_n$, with each $A_i$ the set of actions agent $i$ can take. 
In the most basic case, by treating the environment as static, the \textbf{single-agent Q-learning} algorithm developed by \cite{watkins1992q} gives the optimal policy in an MDP with unknown reward and transition.
\begin{equation}
\begin{aligned}
Q(s,a) & += \alpha [R(s,a) + \gamma V(s') ]  \qquad
V(s) &\leftarrow \max_{a \in A}Q(s,a)
\end{aligned}
\end{equation}
$Q(s,a)$ estimates the value of taking action $a$ when on state $s$, and $V(s)$ the value of the state by taking the best action. 
Extension of this paradigm to multiple agents is difficult. One approach is to assume the environment as passive, each agent with their own reward and transition functions. However, this falsely assumes agent actions do not influence each other~\cite{sen1994learning}. Another approach is to define the $Q-$value function over all agents actions, but introduces a dynamic programming challenge in updating $V$.

In recent years, reinforcement learning has been applied successfully in conjunction with deep learning, using deep neural networks to approximate value functions. A breakthrough comes from \textit{policy-gradient methods}. Traditionally, RL algorithms are \textit{action-value methods}: after learning values of the action, algorithms select actions based on the estimated action values. In contrast, policy-gradient methods learn a parametric policy without consulting the value function~\cite{sutton2018reinforcement}. By policy we mean an agents strategy--- what it does at a given state and time.

Additionally, in cases where the environment is dynamic, it may be optimal to acquire a stochastic policy--- a probability distribution over possible actions. This distribution is updated to associate actions with higher expected rewards with higher probability values. Since probabilities can be over discrete or continuous action spaces, DRL is a useful control framework for negotiation, as the decision to accept or reject an offer is discrete, whereas bidding is on continuous space (on $[0,1]^n$ given $n$ issues). 

\subsection{Policy Gradients}
Call the policy $\pi$ and let parameters $\theta$ define a probability distribution.
The probability of action $a$ is denoted as $\pi(a|s,\theta) = \mathbf{P}\{ A_t = a | S_t = s, \theta_t = \theta  \}$, that is, the probability of taking action $a$ at time $t$ given that the state $s$ and parameters $\theta$.
Similarly, a learned value function, such as using a neural network to approximate the value,  can be represented as $\hat{v}(s,w)$, where $w \in \mathbf{R}^d$ is its weights.

As with action-value RL, policy parameters are optimized to maximize a scalar performance measure $J(\theta)$:
\begin{equation}
J(\theta) = \mathbf{E} [ \sum_{t=0}^{T-1} r_{t+1}   ]
\end{equation}
which describes the expected future aggregate rewards (sum of rewards from $t=0$ until the end). The policy values are updated according to $J$ through gradient ascent:
$$
\theta_{t+1} = \theta_t + \alpha \widehat{J(\theta_t)}
$$
For discrete actions, actions are selected by estimating a \textit{numerical preference value} or \textit{logit} $h(s,a,\theta)$, based on the state, action, and parameter values (weights in a neural net).
Actions are then selected using the softmax distribution:
\begin{equation} \label{eq:softmax}
\pi(a | s,\theta) = \frac{e^{h(s,a,\theta)}}{\Sigma_b e^{h(s,a,\theta)}}
\end{equation}
For instance, for the acceptance strategy, an agent can reject or stop. Associate with these actions $h_R$ and $h_A$ respectively, and a stochastic policy is defined.

However, updating the policy in respect to $J$ requires the \textit{policy-gradient theorem}, which provides guaranteed improvements when updating the policy parameters~\cite{sutton2018reinforcement}. 
The theorem states that change in performance is proportional to the change in the policy, and a full statement is given in Appendix~\ref{sec:policy-gradient-theorem}. The theorem yields a canonical policy-gradient algorithm--- REINFORCE~\cite{sutton2018reinforcement,willianms1988toward,sutton2000policy}. The parameter updates is:

\begin{equation} \label{eq:policy-gradient-update}
\theta_{t+1} = \theta_t + \alpha G_t \frac{\nabla \pi (a_t | s_t,\theta_t)}{\pi (a_t | s_t,\theta_t)} 
\end{equation}
where $G_t$ is the observed reward. Intuitively, the update is the reward multiplied by the gradient of the action probability divided by the action probability. If $G_t$ is high, this increases the chances of visiting that state in the future. Note, the policy gradient is often expressed as  $\nabla \ln \pi (a_t | s_t,\theta_t)$, which yields the fraction through the chain rule.

\subsection{Deep Reinforcement Learning for continuous variables}
Secondly, actor-critic models are useful because they separate the policy space and action space, which means policy selection can occur on a continuous domain. For instance, in a uni-variate control problem, the choice of action can be sampled from a normal distribution. The policy approximation with a normal distribution is:
$$
\pi(a ~|~ s,\theta) = \frac{1}{ \sigma(s,\theta) \sqrt{2\pi} } \exp \bigg( -\frac{(a-\mu(s,\theta))^2}{2\sigma(s,\theta)^2}   \bigg)
$$
During back-propogation, the $\theta$ values are updated such that $\mu$ and $\sigma$ reflect a better reward using the Equation~\ref{eq:policy-gradient-update}.

 \section{Actor-Critic Implementation}\label{sec:actor-critic}
 We have arrived at our main method. Unlike REINFORCE, which only learns a policy, actor-critic models simultaneously learn a value function approximation and a policy. Intuitively, the value function critiques whether an action undertaken by the policy is good, rather than being an absolute measure. Thus, we make a modification to Equation~\ref{eq:policy-gradient-update}, substituting the reward $G_t$ with the value estimate $\hat{q}(s_t,a_t,\vec{w}_t)$ in Equation~\ref{eq:actor-critic-update}:
 
\begin{equation} \label{eq:actor-critic-update}
\theta_{t+1} = \theta_t + \alpha \nabla \ln \pi (a_t | s_t,\theta_t) \hat{q}(s_t,a_t,\vec{w}_t) 
\end{equation}
The process of negotiation thus requires two actor-critic nets--- one for the acceptance strategy and another for the offer strategy. 
The algorithmic procedure is shown in Fig.~\ref{fig:algo-flowchart}, with pseudo-code provided by Algorithm~\ref{algo:training} in the Appendix. We use univariate and single-issue interchangeably, as with multivariate and multi-issue. Next, we describe the architectures of the acceptance and bidding strategy.

 \begin{figure}[!htb]
    \centering
    \includegraphics[width = 1.05\linewidth]{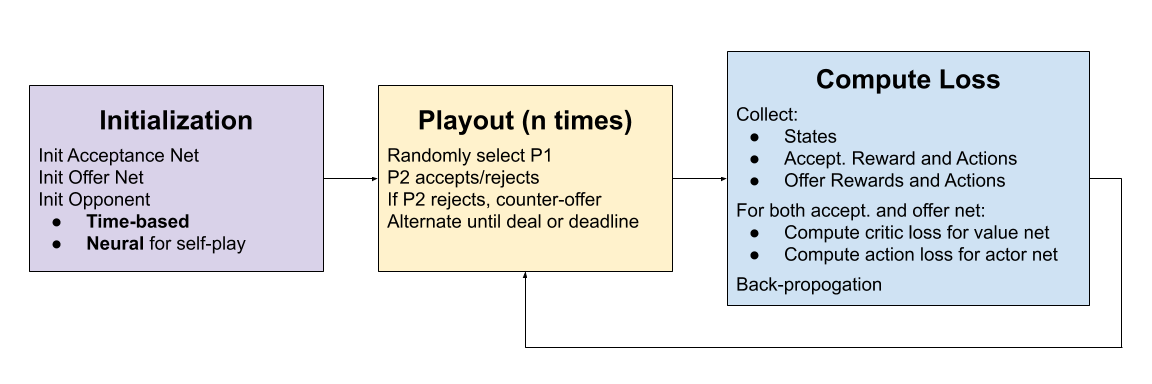}
    \caption{Flowchart of training algorithm.}
    \label{fig:algo-flowchart}
\end{figure}


\subsection{Acceptance Net Architecture}
The first neural network approximates the \textbf{accecptance strategy}. For the univariate case, the input $x$ is a two-element vector consisting of the opponent's offer and the current time step. For the multivariate case, the input is four-dimensional. 

\begin{figure}[!htb]
    \centering
    \includegraphics[width = 1.0\linewidth]{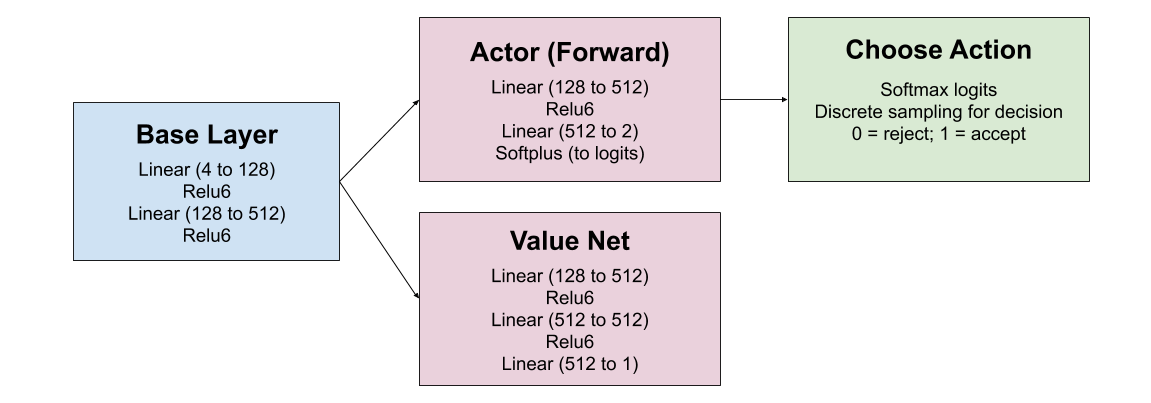}
    \caption{Architecture of the Acceptance Strategy.}
    \label{fig:accept-net}
\end{figure}

At every time step, Accept Net takes in the opponents offer, encodes it to a 512 hidden state using two affine-Relu6 pairs. This base layer is shared between the actor and value network, which facilitates a shared representation~\cite{mnih2016asynchronous}. The actor takes in the embedded state and outputs two logit values, which are softmaxed to choose the appropriate action. Similarly, the value network outputs the expected reward estimate.

The Relu6 is a variant of the Relu functions ($\max \{0,x\})$, but capped at 6. Relu6 layers have been shown to train faster (due to the limit on byte representation) and to encourage the learning of sparse features earlier on~\cite{krizhevsky2010convolutional}. This is important since gameplay is path-dependent and, against a mixed set of opponents, states may be sparse, which we confirmed during preliminary testing. Hyper-parameters were also chosen through testing, reducing layers and the number of hidden states until training behavior changed. The full architecture is shown in Fig.~\ref{fig:accept-net}.

After playout, the critic loss is calculated by taking the mean-squared error (MSE) of the temporal difference--- the difference between the observed rewards and the value network's forward pass. 
Learning parameters are given in Table~\ref{tab:training-params}.

\begin{algorithm}[H] \label{algo:AC-accept}
\SetAlgoLined
\For {Every Reward, State and Action} {
      TDLoss $=$ Reward $-$ $\hat{q}(s_t,a_t,\vec{w}_t)$\;
      $C_{loss}      =($TD Loss \;
      
      LogProbs $=$ $\ln P(a | \pi)$\;
             $A_{loss} = -$ LogProbs $\cdot$ TDLoss\;
      
      Backprop on $C_{loss}$ and $A_{loss}$\;
      }
 \caption{Acceptance Net Actor-Critic Update}
\end{algorithm}

\subsection{Offer Net Architecture}
Next, assuming the agent has rejected the offer, the agent now takes in the same input and decides a counter-offer. Since we have three issues and offers operate on continuous space, the Offer Net must output a vector $o \in [0,1]^3$. To do so, we implement DRL with continuous control, sampling from three different three types of distributions: 1)  multivariate Gaussian, 2) three beta distribution and 3) three Cauchy distributions.

The \textit{multivariate Gaussian} is parametrized by a vector of means $\vec{\mu}$ and covariance matrix $\Sigma$. However, a common assumption in deep learning is that the neural network will capture interdependencies between variables. Hence, an estimate of individual standard deviations along each dimension will suffice. The probability density and policies are given explicitly below.
\begin{equation} \label{eq:multi-variate-gaussian}
\pi(a ~|~ s,\theta) \sim \vec{\mu}(s,\theta), \Sigma(s,\theta)
\qquad
 f_X(x_1, x_2, x_3) = \frac{\exp \big( -\frac{1}{2} (\vec{X}- \vec{\mu})^T \Sigma^{-1} (\vec{X}-\vec{\mu}) \big)   } 
{\sqrt{(\pi}^3 | \Sigma |}   
\end{equation}The \textit{beta distribution} is defined on the $[0,1]$ interval, and defined by two positive shape parameters $\alpha$ and $\beta$. This is useful as offers are held to a finite span. The PDF is:
\begin{equation} \label{eq:beta-distribution}
\frac{x^{\alpha-1}(1-x)^{\beta-1}} {\mathrm {B}(\alpha,\beta)} \qquad \textit{with} \qquad 
   \mathrm {B} (\alpha ,\beta )={\frac {\Gamma (\alpha )\Gamma (\beta )}{\Gamma (\alpha +\beta )}}
\end{equation}\label{eq:beta-distrib}
where $\Gamma$ denotes the gamma function (fractional factorial). Some useful properties of the beta distribution include its intuitive mean and relatively simple expression for variance:
\begin{equation} \label{eq:beta-means}
    \operatorname{E}[X] = \frac{\alpha}{\alpha+\beta}
    \qquad
    \operatorname{var}[X] = \frac{\alpha\beta}{(\alpha+\beta)^2(\alpha+\beta+1)}
\end{equation}{}

Lastly, the \textit{Cauchy distribution} is parametrized similarly to the normal with $\mu$ and $\gamma$ denoted more generally as location and scale. It's density function is given as :

\begin{equation} \label{eq:cauchy-distribution}
\frac{1}{\pi\gamma(s,\theta) \,\left[1 + \left(\frac{x-\mu(s,\theta)}{\gamma(s,\theta)}\right)^2\right]}   
\end{equation}

The neural architecture follows the same actor-critic model described in Section~\ref{sec:actor-critic}. A base layer inputs into the value network and six other neural blocks, estimating the distribution parameters--- three means (locations) and sigmas (scale). For the beta distribution, these are estimates on $\alpha$ and $\beta$. These six variables are then used to sample the offer, which serves as the action used during loss calculation.

\begin{equation}
    \text{Offer} = (o_1, o_2, o_3, t) \sim dist(\theta,s)
\end{equation}

\begin{figure}[!htb]
    \centering
    \includegraphics[width = 1.0\linewidth]{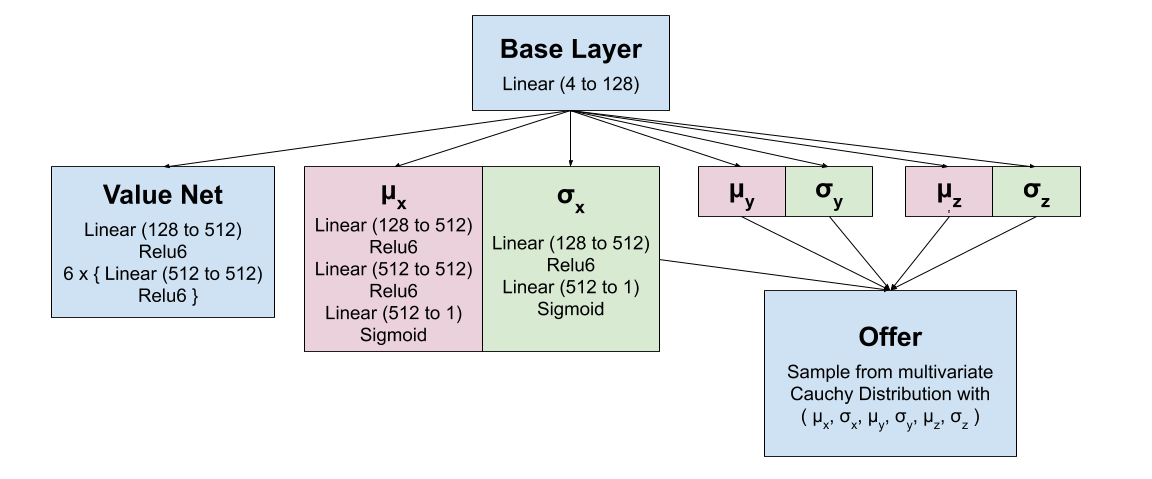}
    \caption{Architecture of the Offer Strategy for multivariate normals and Cauchy distributions. For the beta distribution, the estimated variables were $\alpha$ and $\beta$, and the final forward pass layer is a ReLu layer, instead of a Sigmoid.}
    \label{fig:offer-net}
\end{figure}

The value network consists of seven layers of affine-Relu6 layers. Neural estimates of the mean were conducted with two affine-Relu6 layers, followed with an affine-sigmoid layer to constrain the output between 0 and 1. Sigma estimates used one affine-Relu6 layer and one affine-sigmoid layer. For the beta distribution, the network estimated three pairs of $\alpha$ and $\beta$. Since $\alpha, \beta > 0$, the final sigmoid layer was replaced with a Relu layer. Fig.~\ref{fig:offer-net} shows the architecture in full. Apart from a similar justification for the use of Relus for Accept Net, Relus have documented success for continuous control as well~\cite{lillicrap2015continuous}. Hyper-parameter choice was chosen in a similar way.

Training was undertaken using Adam with learning parameters are given jointly in Table~\ref{tab:training-params}. The exact computations for back-propagation are given in Algorithm`\ref{algo:AC-Offer}. Note, because back-propagation occurs on a continuous domain, log-probabilities of continuous density functions can be positive when variance is small.

\begin{algorithm}[H] \label{algo:AC-Offer}
\SetAlgoLined
\For {Every Reward, State and Action} {
      TDLoss $=$ Reward $-$ $\hat{q}(s_t,a_t,\vec{w}_t)$\;
      $C_{loss} = ($TDLoss $)^2$ \;
      
      LogProbsX $=$ $\ln P(a_X | \pi)$\;
      $AX_{loss} = -$ (LogProbs + Entropy) $\cdot$ TDLoss\;
      Compute $AY_{loss}$, $AZ_{loss}$\;
      Backprop on $C_{loss}$, $AX_{loss}$,  $AY_{loss}$,  $AZ_{loss}$
      }
 \caption{Offer Net Actor-Critic Update}
\end{algorithm}
Here, the entropy is defined as $= 0.5 + \log(2 \pi) \cdot \log( \sigma_Y)$. Adding entropy introduces noise to enable action exploration. Also, high variance means higher loss, so overtime, the variance decreases to improve the precision of the evolved strategy.
 
 \begin{table}[!htb]
 \centering
\begin{tabular}{llll}
\hline
                               & Learning Rate    & Epochs & Optimizer \\ \hline
Accept Net                     & $3e-5$ to $5e-5$ & 8000   & Adam      \\
Offer Net (Gaussian \& Cauchy) & $1e-4$ to $5e-4$ & 4000   & Adam      \\
Offer Net (Beta)               & $1e-3$           & 5000   & Adam      \\
Self-Play                      & $1e-4$           & 3000   & Adam      \\ 
Tit-for-tat                    & $1e-4$           & 5000   & Adam      \\\hline
\end{tabular}
\caption{Deep Learning Training Parameters. Early stopping criteria was convergence in play-out time for 500 epochs. This varied by the concession factor and discount rate} \label{tab:training-params}
\end{table}

\subsection{Reward Scheme}
Accept Net and Offer Net share the same reward scheme. With deadline $T$, value weights $w$, and final offer $x$, the reward given to the neural agent is:

$$
R_t = \begin{cases}
       w^Tx  \qquad & \text{if } t_{f} < T \\
        -K \qquad & \text{if }t_f = T \text{ (conflict deal)}
\end{cases}
$$
This reward function encourages the agent to increase its offer $x$ but not so much it forces a conflict deal and receives low reward.
Unless specified, $-K = -1$.

\subsection{Self-Play: Characterizing Behaviors with Game Theory}
Within bargaining game theory, a focus has been how mechanisms induce norms of fairness, particularly from a branch of game theory called evolutionary game theory (EGT). EGT originates from biology, where it studies the dominance of species through evolutionary pressure, and has been extended to behavioral economics to understand the evolution of behavioral traits.
Nowak et al. showed fairness could be induced if reputation was taken into consideration for the repeated ultimatum game~\cite{nowak2000fairness}, where agents play many one-round negotiations with other agents. 

Their most important finding was that, if agents learned to reject offers they deemed too low, a population of \textit{fair agents would emerge}. Thus, reputation refers to the trait of intentionally reject low offers, and has been confirmed with computational and empirical results~\cite{rand2013evolution}.
We implement a similar study to compare results, using the neural actor-critic model instead of evolutionary methods, and on multi-round negotiation rather than the ultimatum game. Details of implementation are given in Section~\ref{sec:uni-self-play}, where we demonstrate a similar appearance of fairness.

%

\subsection{Against Behavior-Based Agents} \label{sec:methods-t4t}
Lastly, we train our agent against two behavior-based agents. The first is the relative tit-for-tat described in Section~\ref{sec:behavior-based}, and the decision function in Eq.~\ref{eq:t4t-decision-funct}. Furthermore, we implement a Bayesian Tit-for-tat, by estimating the opponent's value weights.
The Bayesian tit-for-tat agent first measures the opponent's concession using its own utility function. Then, it mirrors the amount of concession. Finally, this offer is made as attractive as possible using a Bayesian opponent model~\cite{baarslag2013tit}.

To do this, we first take the ratio of the opponent's offer at $t-\delta$ and $t-\delta-1$ to update the decision utility. If the opponent concedes, then we concede; if they increase their share, we increase ours. Then, we estimate the opponent's utility weights as the mean value of their offers. For instance, if an opponent offers $[1,1,1]$ then $[1,1,0]$, then the utility is estimated as
\begin{equation}
    V_{opp} = [v_x, v_y, v_z] = 6 * \sum_{i=0}^t 
    \frac{[ x, y, z ]}{x+y+z}
\end{equation}
We then implement the Simplex algorithm~\cite{dantzig1955generalized} to maximize this value, fixed upon the decision utility we calculated prior.
While this assumes the opponent makes concessions in particular (preference-based) fashion, it remains a question whether the neural agent can uncover the correct concessions to make.


\chapter{Results}

\section{Theoretical Decision Utilities}
\subsection{Utility Moments}
Before proceeding to results against time-based agents, we first derive the theoretical optimal strategies. Denote the decision utility of the time-based opponent as 
$$U_{opp}(c,t) = P_{res} + (1-P_{res})\big(1-(\frac{t}{T})^\frac{1}{c} \big)$$
This is in essence the same as Equation~\ref{eq:concession-factor}. In our case, $P_{min}$ is the reserve price $P_{res}$ and $P_{max}$ is normalized to $1$. Then our utility is:
\begin{equation} \label{eq:our-util}
\begin{aligned}
U(d)  	&= \bigg( 1-U_{opp}(c,t)  \bigg) d^t 
\end{aligned}
\end{equation}
The maximal point must be one where the marginal utility $\frac{\partial U}{\partial t}$ is 0. Before we take the derivative of $U(d)$, we first take the derivative of $U_{opp}$.
$$
\begin{aligned}
\frac{\partial U_{opp}}{\partial t}  &= \frac{\partial }{\partial t}  P_{res} + (1-P_{res})(1-(\frac{t}{T})^\frac{1}{c}	\\
& = \big(   1- P_{res}  \big) \frac{\partial }{\partial t}   \big(   1-  \big( \frac{t}{T}  \big)^\frac{1}{c} \big) = - \frac{1-P_{res}}{c T ^ \frac{1}{c}}  t^\frac{1-c}{c}
\end{aligned}
$$

We then solve for our marginal utility by the product rule.

\begin{equation}
\begin{aligned}
\frac{\partial U}{\partial t}  &= d^t \ln d (1-U_{opp}) + d^t \bigg(  - \frac{\partial U_{opp}}{\partial t} \bigg) \\
			&=  d^t \bigg(  \ln d (1-U_{opp})   +  \frac{1-P_{res}}{c T ^ \frac{1}{c}}  t^\frac{1-c}{c} \bigg)
\end{aligned}
\end{equation}
Setting the reserve price to $0$ in our experiments, we can derive a much more elegant expression, as $U(c,d,t) = \bigg( \frac{t}{T}  \bigg)^\frac{1}{c} d^t$.
\begin{equation}
\begin{aligned}
\frac{\partial U}{\partial t}  &=  \bigg(   \frac{1}{T^\frac{1}{c}} \bigg) \bigg(   \frac{1}{c} t^\frac{1-c}{c} d^t  +  t^\frac{1}{c} d^t  \ln d \bigg) 
= \frac{t^\frac{1}{c} d^t}{T^\frac{1}{c}} \bigg(  \frac{1}{ct} + \ln d \bigg) 
\end{aligned}
\end{equation}
It is a simple matter to check the second derivative is negative, hence the expression for the condition for the maximal point is 
\begin{equation}
    t = \frac{-1}{c \ln d}
\end{equation}

\begin{figure}[!htb]
    \centering
    \includegraphics[width = 0.7\linewidth]{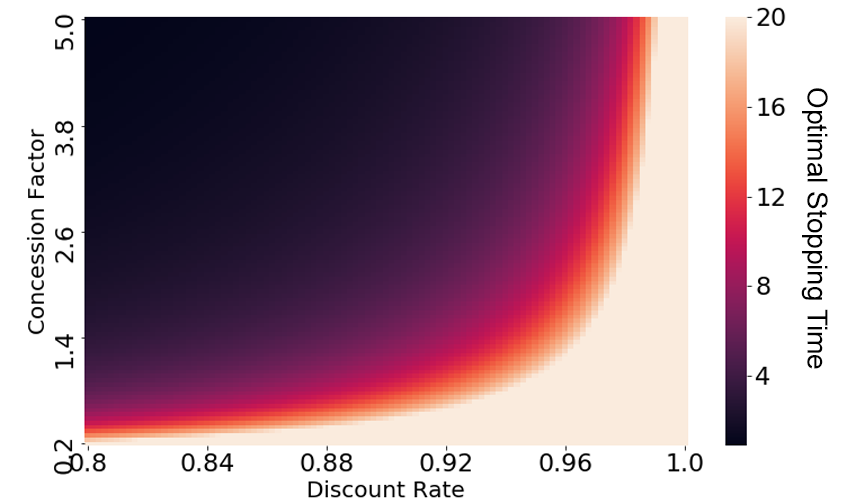}
    \caption{Theoretical optimal stopping time over concession factor and discount rate.}
    \label{fig:theoretical-utilities}
\end{figure}

Interestingly, this value does not depend on the total time $T$ and since $P_{res}$ is a linear transformation of the utility function, this optimal time depends only on the concession factor and discount rate. The optimal stopping time can be expressed as:
\begin{equation} \label{eq:theoretical-stopping-time}
    T_{OPT} = 
    \begin{cases}
        T \qquad \text{if } \frac{-1}{c \ln d} > T \\
        \frac{-1}{c \ln d} \qquad \text{otherwise}
    \end{cases}
\end{equation}
The theoretical values are shown in Fig.~\ref{fig:theoretical-utilities}. A strong phase transition occurs along the $\frac{-1}{c \ln d}$, demarcated by the clearly lighter region.

Additionally, we compute the second derivative of the utility, as it accounts for the error analysis of neural agent in Section~\ref{sec:results-acceptance-strategy}, and the $n$-th moment for generality.
\begin{equation} \label{eq:second-derivative}
\begin{aligned}
    \frac{\partial^2 U}{ \partial t^2} 
    &= \frac{d^t}{T^\frac{1}{c}} \bigg( (\ln d)^2 t^\frac{1}{c} + \frac{\ln d }{c} t^\frac{1-c}{c} 
            + \frac{1}{c}    \bigg( \ln d t^\frac{1-c}{c} + \frac{1-c}{c} t^\frac{1-2c}{c}
    \bigg)      \bigg)  \\
    &=  \frac{d^t}{T^\frac{1}{c}} \bigg( (\ln d)^2 t^\frac{1}{c} + \frac{2 \ln d }{c} t^\frac{1-c}{c} + \frac{1-c}{c^2} t^\frac{1-2c}{c}   \bigg)  
\end{aligned}
\end{equation}
\begin{equation}
\frac{\partial^n U}{\partial t^n} = \frac{d^t}{T^\frac{1}{c}} \sum_{i=0}^n \binom{n}{i} (\ln d)^{n-i} \frac{1}{c^i} 
\bigg(  \prod_{j=1}^i 1-(j-1)c  \bigg) t^\frac{1-ic}{c}
\end{equation}


\begin{figure}[!htb]
    \centering
    \includegraphics[width = 1.0\linewidth]{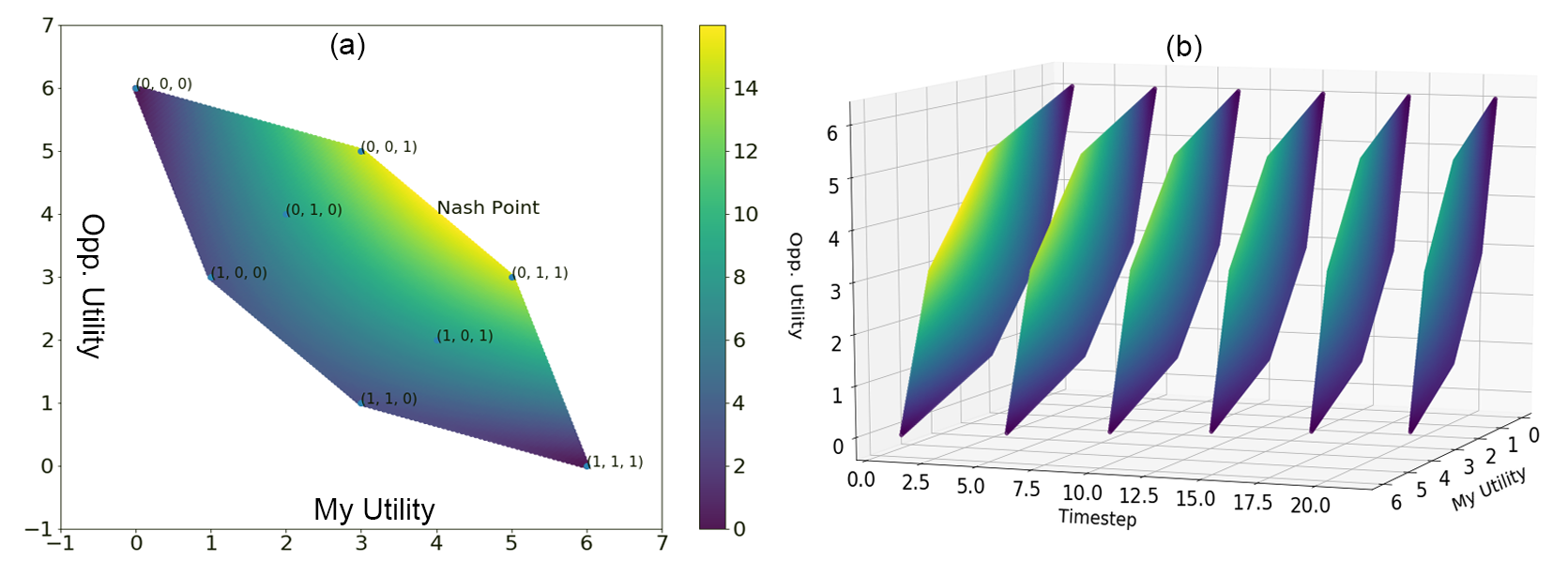}
\caption{Outcome space of Negotiation. Fig.~\ref{fig:outcome-space}a) shows the outcome space at $t=1$. The vertices of the outcome space polytope is mapped from the vertices of the action space, at $(0,0,0), (0,0,1), (0,1,1), (1,1,1)$. Fig.~\ref{fig:outcome-space}b) shows the evolution of the polytope over time, with a discount rate of $0.9$. 
The decision to counter-offer is made by the subsequent time-step, rather than the current. 
The colors denote the Nash Product, with the Nash solution lying at $(4,4)$.}
\label{fig:outcome-space}
\end{figure}

\subsection{Outcome Space}
In the field of automated negotiation, preferences are typically visualized through an \textit{outcome space plot}. The axes are utilities of Player $A$ and $B$. Possible outcomes $\omega \in \Omega$ are mapped to $(u_A(\omega), u_B(\omega))$. 
Fig.~\ref{fig:outcome-space} shows this plot for our negotiation process. In a), the Pareto frontier is shown by the right-most edges of the polytope.

By theorems of fixed points and the simplex algorithm~\cite{wong2015bridging, dantzig1955generalized}, the vertices in the outcome space must come from the vertices in the action space $[0,1]^3$. The action space vertices that outline the frontier are found to be $(0,0,0), (0,0,1), (0,1,1), (1,1,1)$. Intuitively, these are points that offer the greatest marginal utilities to P1 and P2, based on their value weights $w_A = (1,2,3)$ and $w_B = (3,2,1)$. 
The piece-wise equation for the Pareto Frontier $PF(U_A, U_B)$ is given as follows, in Equation~\ref{eq:pareto-frontier}:
\begin{equation} \label{eq:pareto-frontier}
    PF(U_A, U_B) = \begin{cases}
    U_B + \frac{1}{3}U_A - 6 = 0    \qquad & \text{for } 0 \geq U_A \geq 3 \\
    U_B + U_A - 8 = 0               \qquad & \text{for } 3 \geq U_A \geq 5 \\
    U_B + 3U_A - 18 = 0             \qquad & \text{for } 5 \geq U_A \geq 1 
    \end{cases}
\end{equation}
This equation allows us to calculate the bid distribution and determine the \textit{efficiency} of an agent's bid strategy, provided in Section~\ref{sec:point-2-line} in the Appendix. 
Furthermore, the Nash Solution ($\max U_A U_B )$ lies at $(4,4)$ given by the offer $(0,0.5,1)$, provides a benchmark for the \textit{fairness} when playing against behavior-based agents.

\section{Acceptance Strategy} \label{sec:results-acceptance-strategy}
\subsection{Behavioral dynamics: Cliff-walking vs optimal play}
The central question for an acceptance strategy is when given an offer, whether or not to accept or wait for potentially better future offers. However, if the agent fails to accept before the deadline, then the conflict deal is enacted and both agents do not receive any reward. Given a discount rate and opponent concession factor, the goal is to find the best moment to accept an offer, inferring from their prior offers.

\begin{figure}[!htb]
    \centering
    \includegraphics[width = 0.8\linewidth]{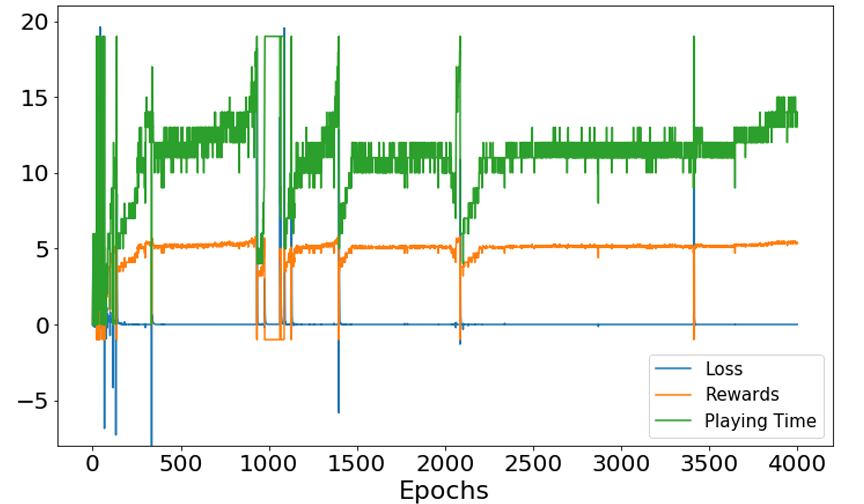}
    \caption{Loss, rewards and total time of the DRL agent training against a time-based agent with $c = 1.0$. After a few epochs of random search, the agent learns that increased playtime comes with greater reward. However, as this time increase to the deadline,the reward drops sharply. 
    }
    \label{fig:multivariate-training}
\end{figure}

Thus, the acceptance strategy can be seen as an optimal stopping problem with an additional cliff-walking problem to solve. Fig.~\ref{fig:multivariate-training} shows the loss, rewards, and playing time as the network trains against a linear agent ($c = 1.0$) with no discount ($d = 1.0$). Through stochastic sampling of new points, the agent notices greater reward by waiting, illustrated by gradual trends in playing time (green). However, once the agent reaches the deadline at 20 rounds, the conflict deal is enacted and a reward of $-1$ is issued, producing a large loss.
We present only the multivariate case, as results for the univariate case are the same but with lower complexity.

\begin{figure}[!htb]
    \centering
    \includegraphics[width = 1.0\linewidth]{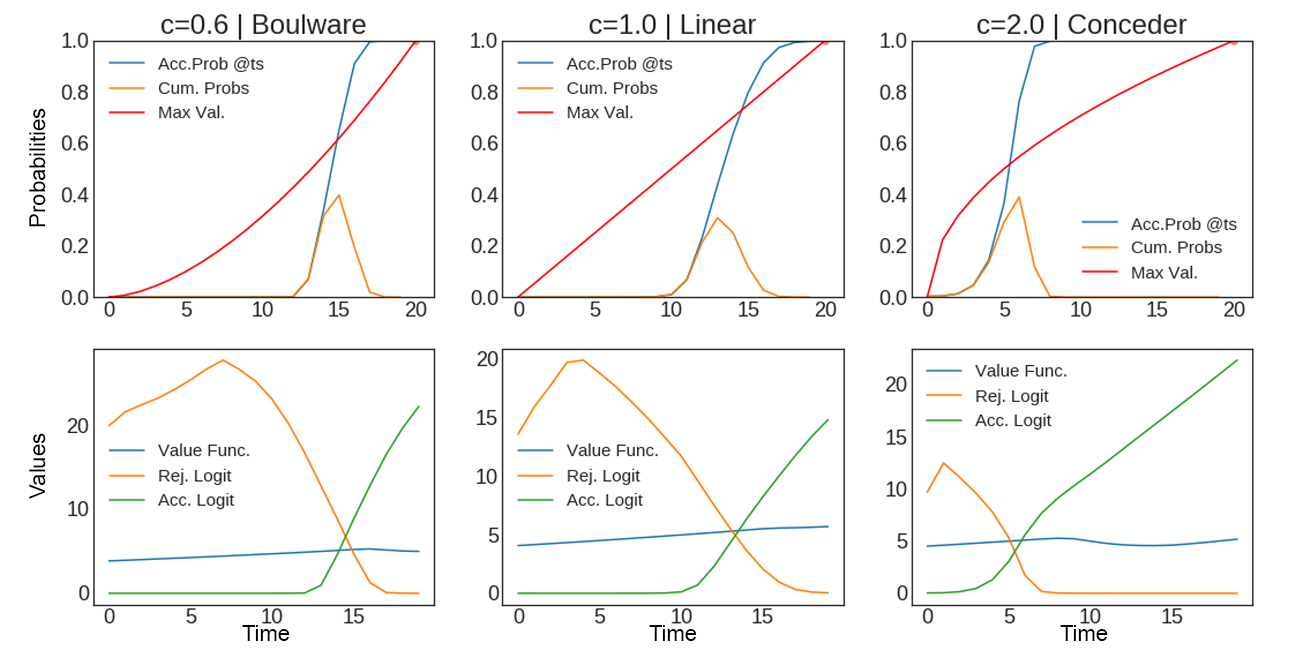}
    \caption{Evolution of acceptance probabilities and logit values without discounting ($d = 1.0$). Row 1 shows the acceptance probabilities at each time step (blue), the cumulative probability of first success (orange), and max value at each time step. 
    The pink point denotes the optimal value (Eq.~\ref{eq:theoretical-stopping-time}). 
    As the concession factor $c$ increases, stopping time decreases. Row 2 shows where the logits "change places," corresponding to where the cumulative probability is maximal.
    }
    \label{fig:multivariate-accept-no-discount}
\end{figure}

To analyze the stopping time, we consider the evolution of acceptance probabilities during gameplay against time-based opponents. Fig.~\ref{fig:multivariate-accept-no-discount} shows the logit values used in Eq.~\ref{eq:softmax} and acceptance probabilities against Boulware, Linear, and Conceder agents.

The first row shows the acceptance probabilities at each time step. The cumulative probability (orange) denotes the likelihood the game ends at a certain time step, given as:
\begin{equation}
    P_{cum} (t) = P_{t}(\text{Accept}) \prod_{i=0}^{t-1} P_{i} (\text{Reject} )
\end{equation}
Since the discount rate is $1.0$, the optimal value is waiting until the final point in time. 
The decrease in stoppage time shown by the right-shifting cumulative probability is sub-optimal, although this is not uncommon in conservative agents. In the value function (blue, second row), there is also a slight decrease after the logit values cross. This indicates that the expected reward at these times may be the same.

\begin{figure}[!htb]
    \centering
    \includegraphics[width = 1.0\linewidth]{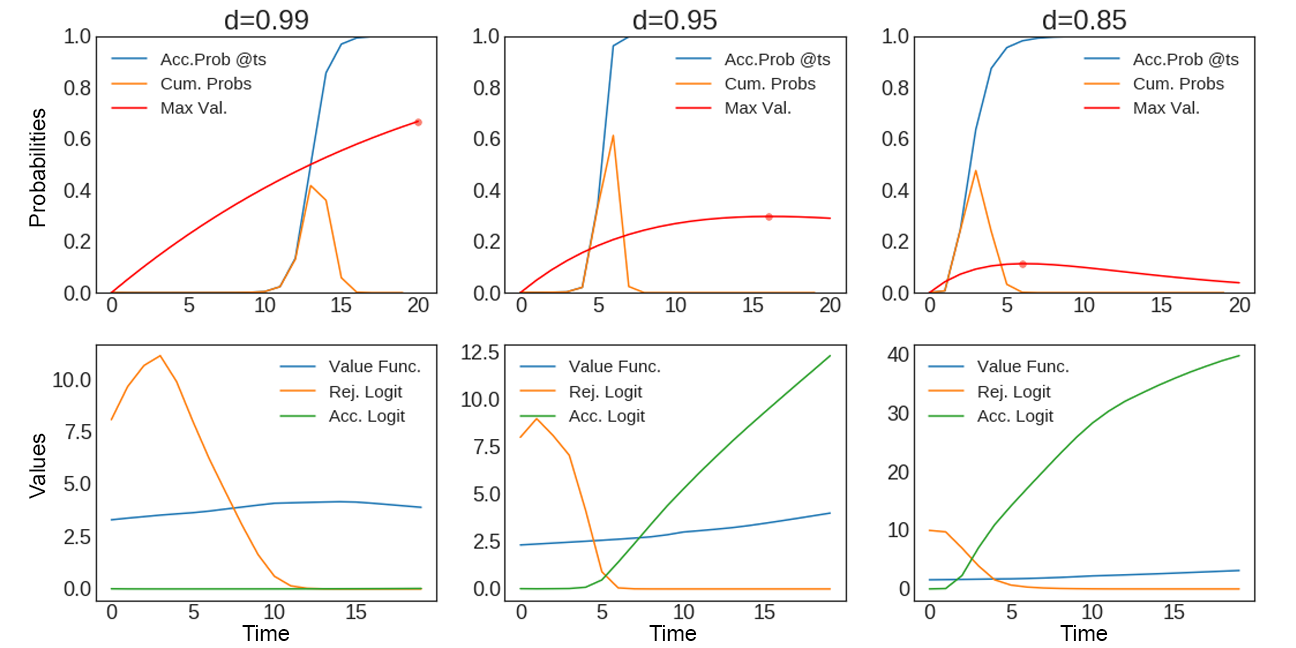}
    \caption{Evolution of acceptance probabilities and logit values with discount. Row one shows the acceptance probabilities at each time step (blue), the cumulative probability of the first success (orange), and the pink point denotes the theoretical maximum prescribed by Equation~\ref{eq:theoretical-stopping-time}. As the discount rate increases, the optimal maximum and stoppage time decreases.}
    \label{fig:multivariate-accept-discount}
\end{figure}

Another way to see this is to consider the marginal utility over time. Since Boulware agents only concede towards the end, the Neural agent is forced to wait to achieve comparable results, whereas it may be ``satisfied" earlier against Conceders. The marginal utility of the Boulware agent is thus much greater towards the end, whereas marginal utility is high at the beginning against Conceders. Explicitly, for $U = 1 - 1 + \big( \frac{t}{T} \big)^\frac{1}{c} = \big( \frac{t}{T} \big)^\frac{1}{c}$, the expression for marginal utility is:

\begin{equation}
    \begin{aligned}
    \frac{\partial U(c)}{\partial t} &= \frac{1}{cT^\frac{1}{c}} t^\frac{1-c}{c}
    \end{aligned}
\end{equation}
This analysis is corroborated further once we introduce the discount rate. Fig.~\ref{fig:multivariate-accept-discount} shows the acceptance probabilities and logits once discounting is introduced. 
Looking at the red curve, $d=0.99$ has positive marginal utility and for $d=0.85$, the marginal utility is negative from time step 7 onwards. For $d=0.95$, the utility function is relatively flat after time step 10, which means the marginal utility is close to 0. Here, we observe the greatest time deviation.

\subsection{Marginal Analysis: Marginal Utility determines Error}
\begin{figure}[!htb]
    \centering
    \includegraphics[width = 1.05\linewidth]{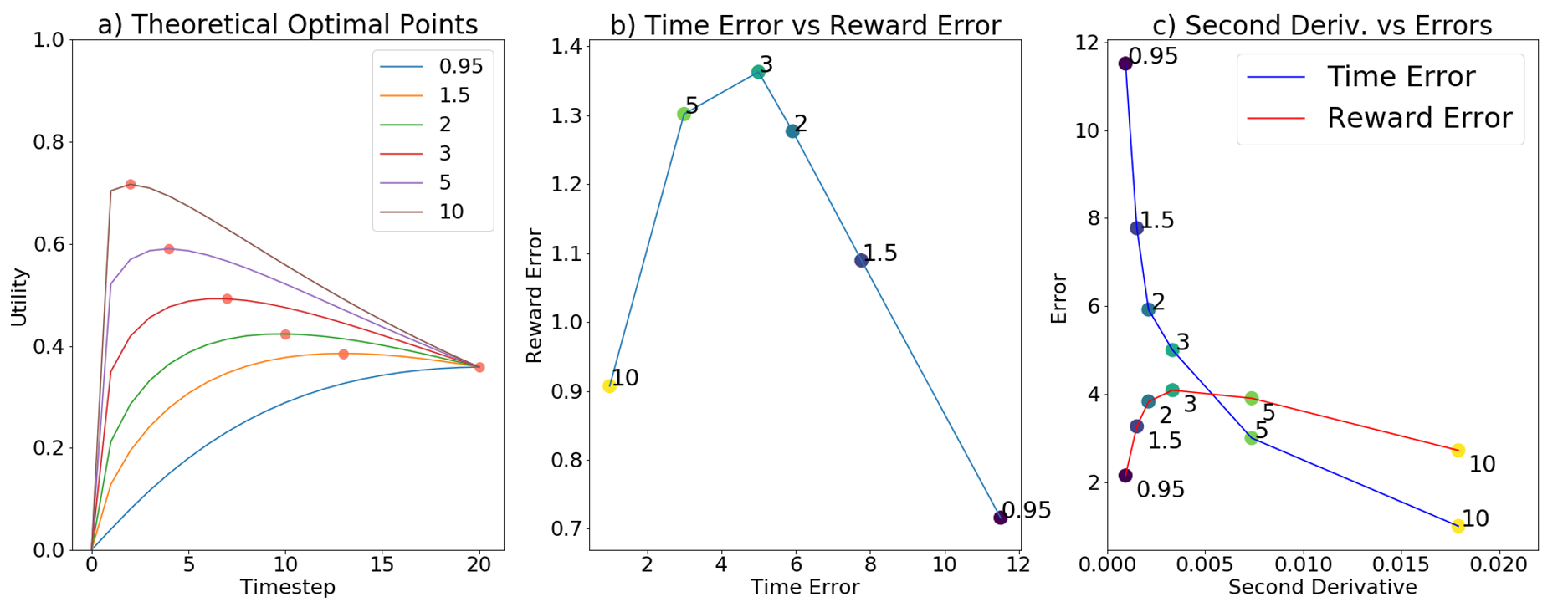}
    \caption{Optimals vs second derivative from 100 gameplays. Rew. error scaled by 3.}
    \label{fig:second-deriv}
\end{figure}

Due to the stochastic nature of deep learning, it's difficult to construct a precise mathematical proof of how changes in marginal utilities push against each other. However, we can test this empirically. The neural agent played against a set of different agents, with concession factors of 0.95, 1.5, 2, 3, 5, 10. In Fig.~\ref{fig:second-deriv}a), the curves show our max utility (Equation~\ref{eq:our-util}) and the red dot shows the optimal stopping time given by Equation~\ref{eq:theoretical-stopping-time}. Note, as $c$ increases, the curves grow sharper and since $c>1$, the magnitude of the second derivative strictly increases. 

Fig.~\ref{fig:second-deriv}b) shows an inverse relationship between the time error and the reward error. The ``peakier" the curve, the more likely the Neural net selects the optimal time. However, deferral by even one time step leads to large amounts of diminished utility, hence creating the larger reward error. In contrast, using the second derivative derived in Equation~\ref{eq:second-derivative}, we observe in Fig.~\ref{fig:second-deriv}c) that as the second derivative approaches 0, the time error increases. 

Having shown what produces the reward and time errors, we can address our sub-problem about limitations. For future work, we may dynamically reduce the learning rate using the second derivative and distance to the deadline for better convergence. Numerical results are summarized in Table~\ref{tab:marginal-utils}.

\begin{table}[!htb]
\begin{tabular}{llllllll}
\hline
$c$                        & $0.3$   & $0.95$ & $1.5$  & $2$    & $3$    & $5$    & $10$   \\ \hline
Time Error                 & 3.95    & 11.62  & 7.81   & 5.86   & 4.99   & 3.0    & \textbf{1.0}    \\
Reward Error               & \textbf{-0.0234} & -0.712 & -1.054 & -1.278 & -1.366 & -1.302 & -0.907 \\
Second Deriv. ($x100$) & 0.0722  & -0.094 & -0.152 & -0.223 & -0.389 & -0.777 & -1.886 \\ \hline
\end{tabular}
\caption{Tabular Results of 100 gameplays sat different concessions.} \label{tab:marginal-utils}
\end{table}


\subsection{Preference-based concessions produce fairer outcomes}
\begin{figure}[!htb]
    \centering
    \includegraphics[width = 0.5\linewidth]{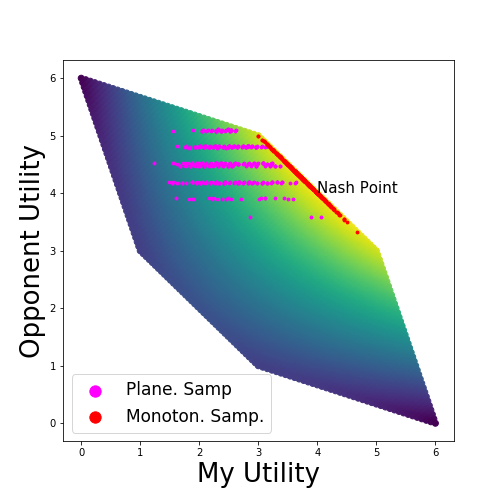}
    \caption{Distribution of final, accepted offers ($c=1.0$, $d=0.94$). The preference-based bidding (red) produces offers on the Pareto Frontier.
    Planar sampling (magenta) occur in intervals as the time-based agent samples points based on $u_B(t)$.}
    \label{fig:accept-monotoni-vs-random}
\end{figure}

Finally, we consider optimality. Since the final offers depend on the time-based agent, so do the optimality measures. Thus, the way opponent agent algorithmically constructs their offers will appear differently in the outcome space. Fig.~\ref{fig:accept-monotoni-vs-random} shows the distribution of accepted offers after 400 gameplays, with $c =1.0$ and $d = 0.94$. The first randomly samples from the plane that satisfies the following condition:
\begin{equation} \label{eq:time-decision-plane}
U_d(t) = w^T X = w_1x_1 + w_2 x_2 + w_3x_3
\end{equation}
where $U_d$ is the decision utility at time $t$ and $w_i$ is the weighed utility for issue $x_i$. The second uses a \textit{preference-based, monotonic concession} strategy--- it satisfies Equation~\ref{eq:time-decision-plane}, but concedes starting from the issue it values the least ($\min\{ w_1, w_2, w_3 \}$). Multivariate Gaussian noise with a standard deviation of $0.05$ is added to prevent deterministic offers.
When the time-based agent uses preference-based, monotonic concession strategies then this guarantees offers to lie on the Pareto Frontier.

\begin{table}[!htb]
\begin{tabular}{llllll}  
\hline & $d_{Nash}$ & $BD(\Omega)$ ($d_{Pareto}$) & Av. Reward & Av. Time \\ \hline
Planar Samp. & 1.64 & 0.54 & 1.84 & 4.89 \\
Preference-based Concession & \textbf{0.46} & \textbf{0.00} & \textbf{2.61} & \textbf{5.7} \\
Pure Random & 1.92 & 1.24 & N/A & N/A \\ 
\hline
\end{tabular}
\caption{Sampling Results}\label{tab:offer-measures}
\end{table}

 The preference-based method produces offers that lie on the Pareto Frontier. Because of this optimality, the neural agents play on average a longer time when its opponent follows this strategy. Random planar sampling yields considerably better results than pure random sampling, with a bid distribution difference of $0.7$ (shown in Table~\ref{tab:offer-measures}). The magenta points in Fig.~\ref{fig:accept-monotoni-vs-random} arise because, for every time step, the decision utility is fixed for fixed $c$. Table~\ref{tab:offer-measures} summarizes the mean outcomes of the gameplays, with the preference-based concession performing the best.

\section{Bidding Strategy}
\label{sec:results-offer-strategy}
\subsection{Precision in Single-Issue Negotiation}
\begin{figure}[!htb]
    \centering
    \includegraphics[width = 1.0\linewidth]{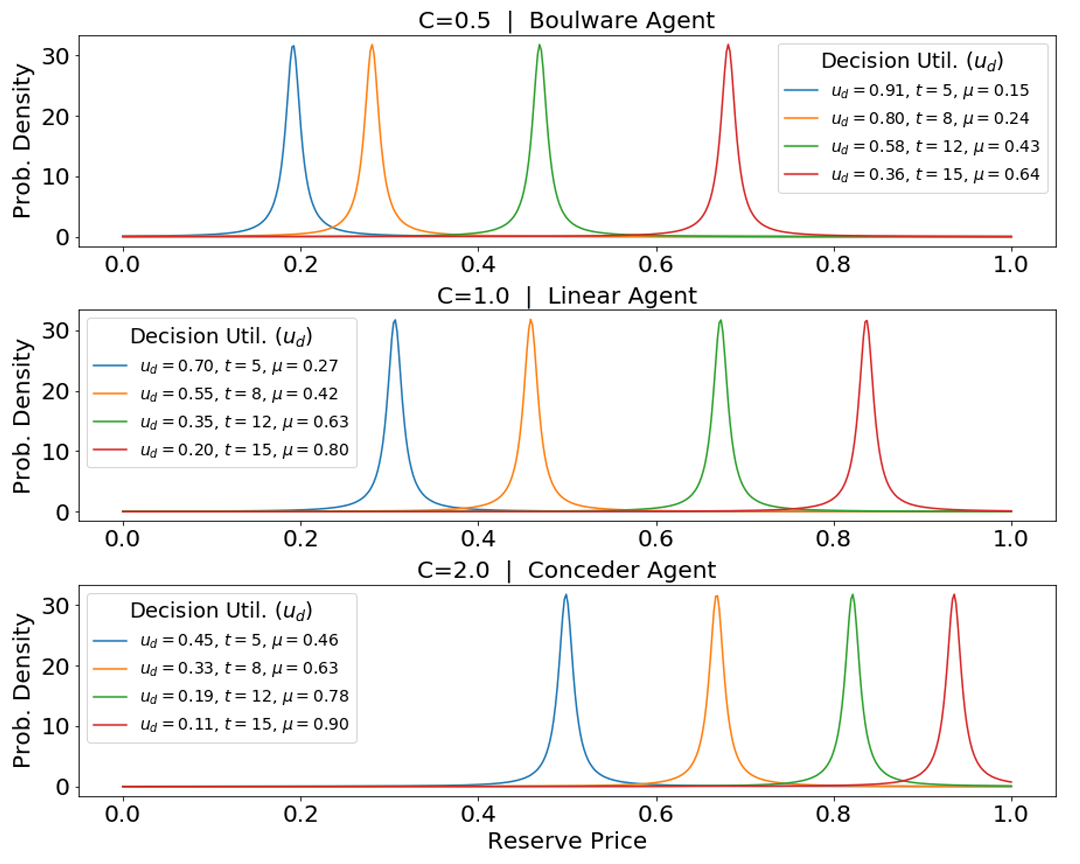}
    \caption{Cauchy distributions based on concession factor and decision utility. When $c$ is low (Boulware), Cauchy means are clustered tightly for low $t$, spread out for high $t$. When $c$ is high (Conceder), values are clustered for high $t$ and spread out for low $t$. }
    \label{fig:uni-cauchy}
\end{figure}
Before evaluating performance on the multivariate case, we verify the univariate case. Fig.~\ref{fig:uni-cauchy} shows the action policies given by Cauchy distributions for specific decision utilities. 
As the concession factor increases, the distribution of means transfers from right- to left-skew. This can be attributed to the magnitude of the marginal utility. Cauchy means are clustered tightly for the Boulware agent when $t$ is low, as the marginal utility is low early on. However, as time passes and the Boulware agent begins to concede greatly, the distance between means increase. Conversely, when the opponent is a Conceder, concession begins early, so marginal utility is large when $t$ is small, leading to right-skew. The Conceder case is not as pronounced as the Boulware case due to the cliff at the deadline. As expected, means are spaced out linearly against linear agents. 

In sum, the change in decision utility affects the distribution of the means. For completeness, Fig.~\ref{fig:cauchy-heat} in the Appendix shows a heat map of how the neural agent's utility changes in respect to the opponent's.

\subsection{Multivariate Training Dynamics}
\begin{figure}[!htb]
    \centering
    \includegraphics[width =0.8\linewidth]{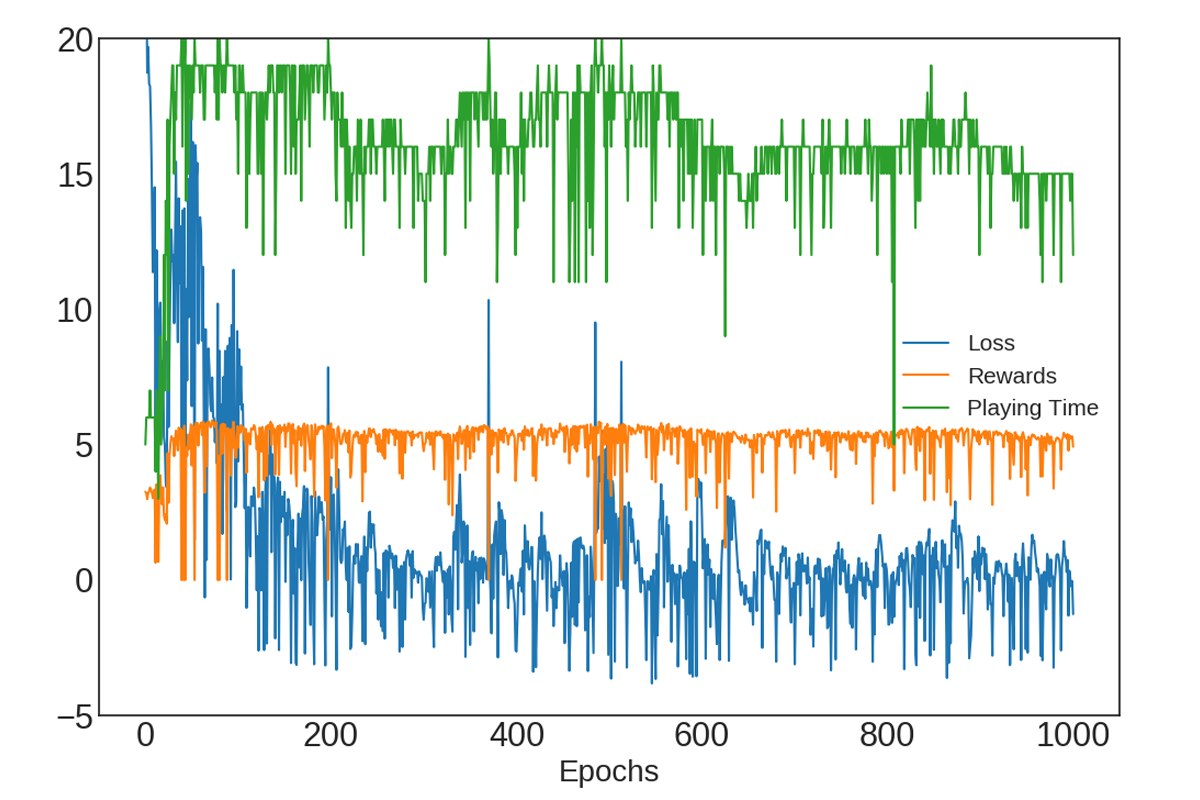}
    \caption{Training statistics of the Cauchy Offer Net. The neural agent learns to induce rejection from the time-based agent so negotiation ends near the deadline.}
    \label{fig:multi-offer-training}
\end{figure}

Next, we present the multivariate case. We trained on a grid of concession factors for fixed discount rates. We denote the three issues as issues $X$, $Y$, and $Z$.
Fig.~\ref{fig:multi-offer-training} shows the first 1000 epochs, using a multivariate Cauchy distribution with $c=0.3$ and no discount. Unlike training Accept Net, cliff-walking is less present, as the final action (accept) lies with the opponent and time-based agents are very likely to accept.
The agent quickly learns to wait longer, converging at a higher time step. 

\begin{figure}[!htb]
    \centering
    \includegraphics[width =1.0\linewidth]{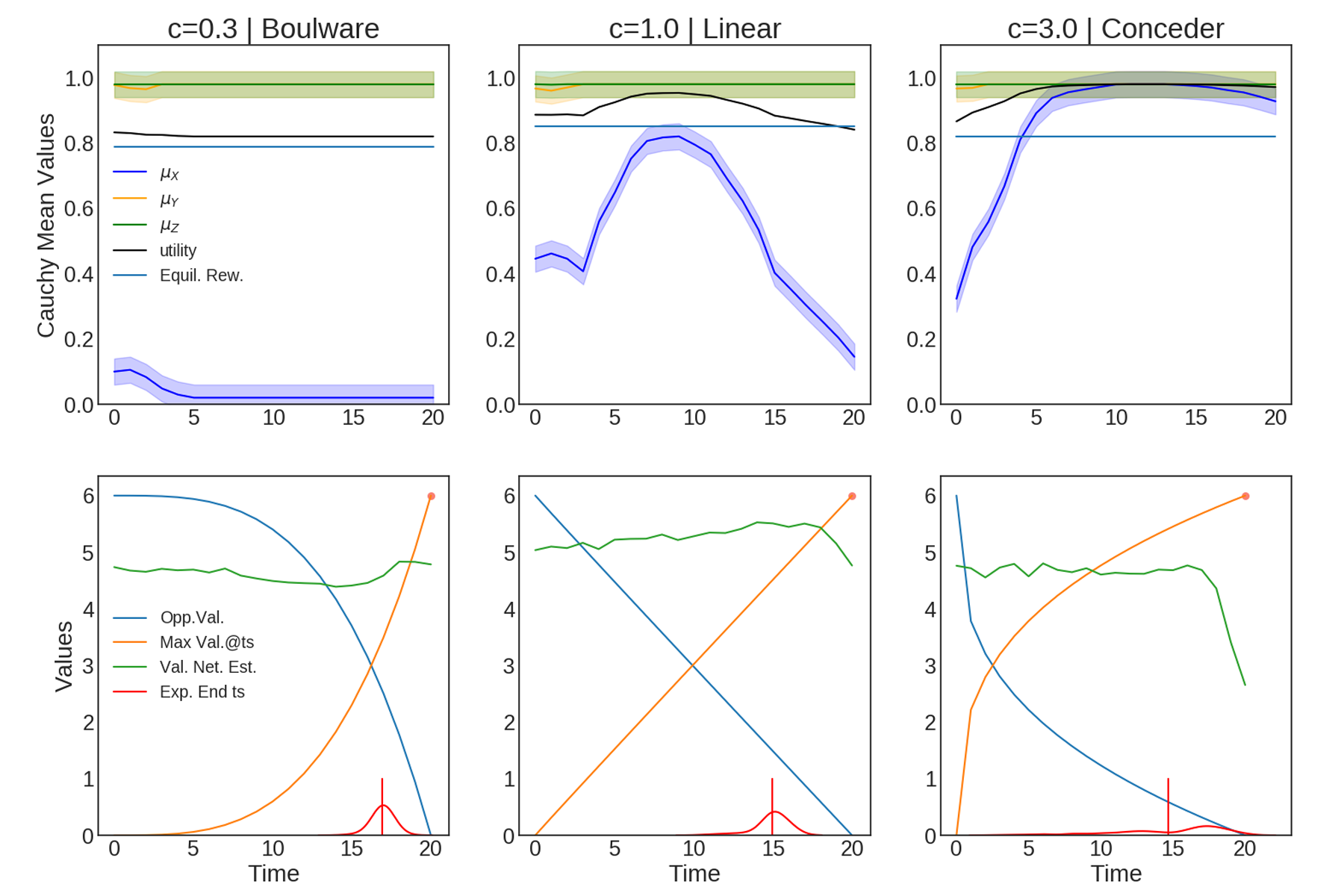}
    \caption{Variation of Cauchy means over time with no discount rate. The expected reward hovers around 84 percent of the full utility. All stoppage times are high, but still decreases as the concession factor increases due to reduced marginal utility. }
    \label{fig:multi-offer-no-discount}
\end{figure}

However, unlike accept net, this is not simply a binary action where rejecting an offer leads to the next round. The agent has to produce offers that induce rejection from the time-based agent. Fig.~\ref{fig:multi-offer-no-discount} shows how Cauchy means vary over time. The first row shows the individual Cauchy means for issues $X$, $Y$ and $Z$ respectively. Shown in light blue is also the normalized utility (divided by the total possible utility of 6) and the equilibrium payout from 100 gameplays. The second row shows the opponent's decision function (blue) and our maximum utility at each time step. The value estimate given by the value net is shown in green. At the bottom, the red line shows the mean stoppage time, with the distribution of times shown with a kernel density estimate.

Since the cliff-walking aspect is not as prominent (the only case where the conflict deal is enacted is if the agent proposes the full amount at the end), all stoppage times are relatively high, although diminishing stoppage time is still observed when the marginal utility is lower. For instance, when $c = 1$, the marginal utility is constantly $\frac{1}{20}$ and the second derivative is $0$, the stoppage time is $15.1$. Comparably, when $c=0.3$, the marginal utility is increasing and the mean stoppage time is $16.9$.

Note issue $X$ varies the most, either through concession (Fig.~\ref{fig:multi-offer-no-discount}a)) or increase of the offer value as shown in  Fig.~\ref{fig:multi-offer-no-discount}c). At a glance, this may be counter-intuitive, since a change in $Y$ or $Z$ would yield the most marginal gains for the agent. However, the opponent values issue $X$ the most, which means $X$ produces the largest amount of gradient for the least amount of loss during concession. As a result, the agent learns the negotiate close or along the Pareto Frontier, which we show later in the distributional analysis. Secondly, there is a clear progression between Boulware and Conceder strategies when comparing the linear agent to the Boulware agent.

Additionally, in the bottom row, the green value function remains fairly constant throughout, until the drop off towards the end induced by the deadline. The value function remains flat since the expected value is constant--- so long as the neural agent sticks to its strategy, the payout will not change. While performance is not optimal, it achieves more than $80\%$ of the optimal which is typical of risk-averse agents whose behaviors are generally conservative on estimates~\cite{sandholm1999bargaining}. 

\begin{figure}[!htb]
    \centering
    \includegraphics[width =1.0\linewidth]{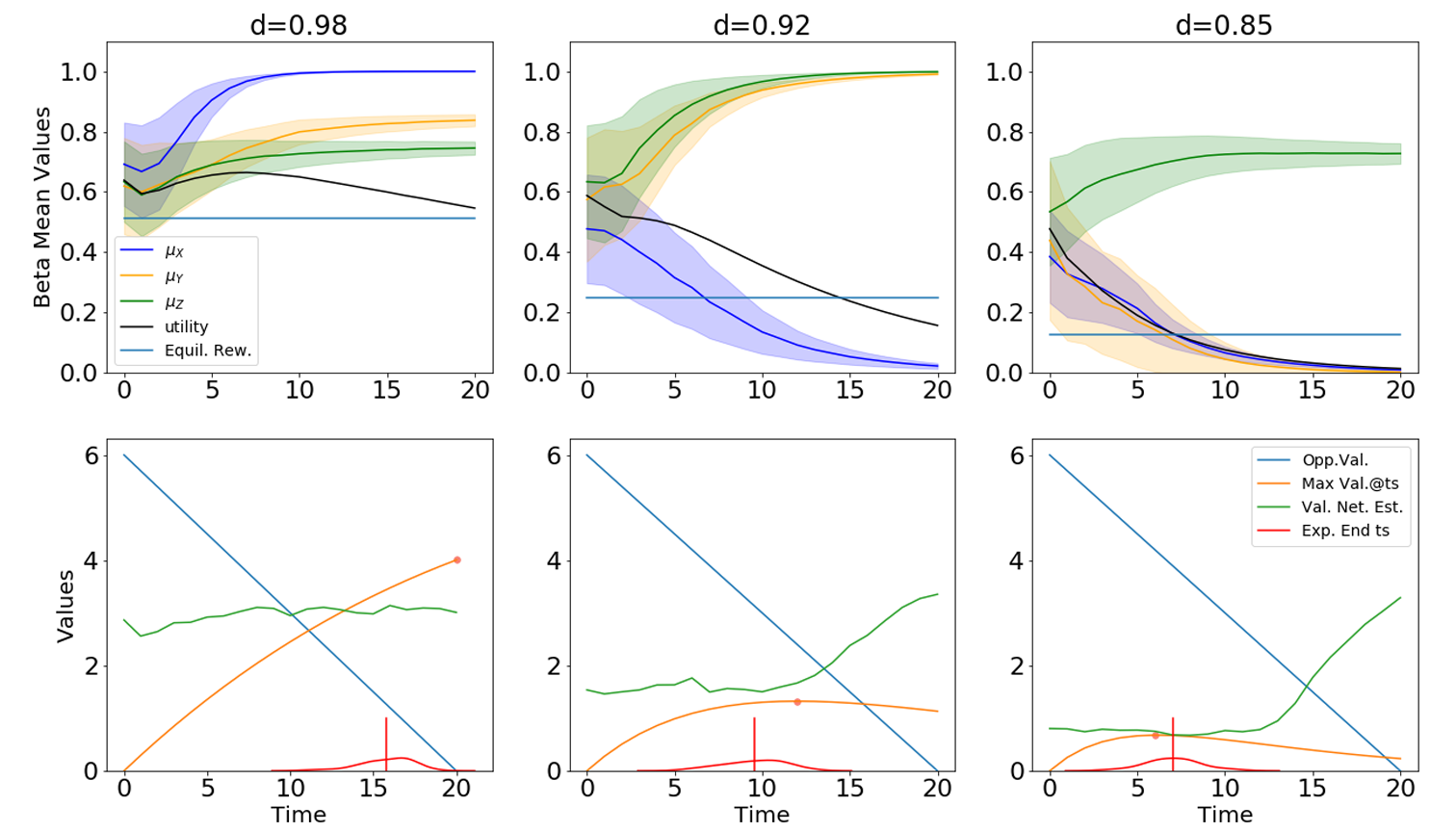}
    \caption{Variation of beta distribution means over time with discount rate. Each mean is calculated using Equation~\ref{eq:beta-means}. Gameplay results fall consistently within $10\%$ of the expected optimal stoppage time. Values begin around the same region with high variance as the agent is unsure what opponent it is playing against, but decreases with time as certainty towards its opponent's strategy grows. $c = 1$. }
    \label{fig:multi-offer-discount}
\end{figure}

Next, we compare this to the case when discounting is introduced, using the beta distribution as an example. Fig.~\ref{fig:multi-offer-training} shows the gameplays of a neural agent using the beta-distribution. The first row shows the evolution of the multivariate distribution means (blue, orange and green for $X$, $Y$, and $Z$ respectively), the evolution of the normalized utility (black), and the reward under 100 gameplays. The second row shows the theoretical maximums (orange) and mean stoppage time (red). Immediately, we observe that the agent begins sampling around the same initial values, then alters its strategy as it learns more about the opponent. The way it alters its strategy varies depending on its own inherent discount rate, which demonstrates \textit{adaptability}.

Finally, issue $X$ is again the issue with the most variation, with the same argument that its change produces the highest gradients during gameplay, due to the opponent valuing $X$ the most. Mean stoppage time is close to the optimal, with $8\%$ mean deviation. 
While this is not precisely optimal, it is quite good. To understand what causes this limitation, we analyze the probability distributions.

\subsection{Offer Strategy requires Sensitivity to Variance} \label{sec:multi-var-distrib-comparison}
\begin{figure}[!htb]
    \centering
    \includegraphics[width =1.0\linewidth]{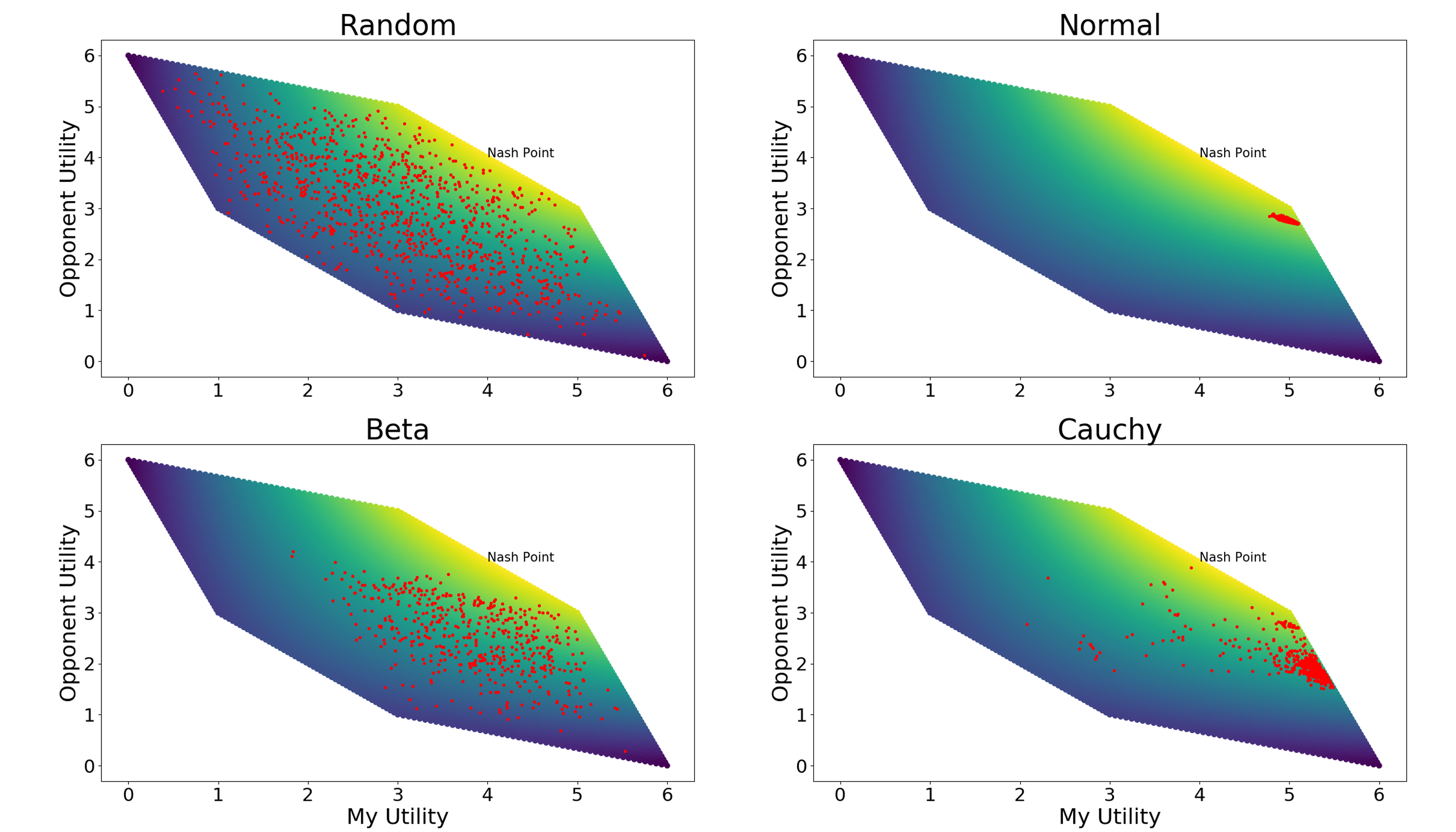}
    \caption{Distribution of outcomes based on different sampling distributions. Neural agents were trained on 3000 games, then played against a linear agent. The random agent plays.
    Statistics are summarized in Table~\ref{tab:offer-results}}
    \label{fig:multi-offer-nash}
\end{figure}

We compare the outcome space of agents using Gaussian, Cauchy, and beta distributions, after playing 3000 rounds against batches of mixed opponents. Fig.~\ref{fig:multi-offer-nash} shows the distribution of final offers given by each agent, with the addition of a random agent, when playing against a linear agent with no discount rate. Since time-based agents make monotonic concessions, lower y-values imply longer gameplay times.

As expected, the random agent produces offers distributed randomly in the outcome space. 
At a glance, the beta distribution outcomes bear the most resemblance to the random agent. This is due to the beta distribution's initial high variance. This can be adjusted by increasing the constant added to the initial values of $\alpha$ and $\beta$. However, also note that the scattered points are on average greater than 3.

The normal distribution produces the most consistent results, with results clustered around the $(0,1,1)$ vertex. The Cauchy distribution performs similarly, but on average performs better, with a maximum value of $5.5$ and average $5.3$. However, it also has a much greater variance when compared to the Normal distribution. We can conclude convergence to optimal play requires sufficient initial variance to prevent convergence to local optima.
Additionally, the maximum value achieved by any of these distributions was achieved by the beta distribution. These values are presented in Table~\ref{tab:offer-results}.

\begin{table}[!htb]
\begin{tabular}{llllll}
\hline
 & $d_{Nash}$ & $BD(\Omega)$ & Av. Reward & Av. Time & Reward. Range \\ \hline
Rand. Samp. & 1.901 & 1.246 & NA & NA & NA \\
Beta & 1.741 & 1.0018 & 3.587 & 10.11 & 5.378 \\
Normal & \textbf{1.585} & \textbf{0.0815} & 4.993 & 11.03 & \textbf{0.294} \\
Cauchy & 2.261 & 0.218 & \textbf{5.051} & \textbf{14.66} & 4.403 \\ \hline
\end{tabular}
\caption{Gameplay outcomes from 400 games against linear agent. Sampling through the normal distribution givese the fairest and most consistent results, whereas the Cauchy provides the highest expected reward.} \label{tab:offer-results}
\end{table}

The normal distribution produces consistent results, with the lowest bid distribution and reward range, and also is closest to the Nash Solution. In expectation, the Cauchy distribution produces better results but with a higher bid distribution and is farther from the Nash Solution.
However, having a large distance from the Nash Point is not necessarily bad, as exploiting the opponent's strategy leads to higher rewards. Without discount, the neural agent can improve its own outcomes by waiting.

The four panels in Fig.~\ref{fig:multi-offer-nash} reveal two opposing forces that make continuous DRL difficult in this domain. Convergence to optima requires high variance, yet avoiding the conflict deal requires low variance to prevent sampling the conflict deal. 
For further proof, consider that the time error decreases with discount rate, comparing Fig.~\ref{fig:multi-offer-no-discount} and Fig.~\ref{fig:multi-offer-discount}. The discount rate shifts the optimal away from the cliff, thus sampling around the optimal produces less error due to slow change in marginal utility, and the smooth reward function allows more accurate function approximation~\cite{lillicrap2015continuous,sutton2018reinforcement}.

The next step is to compare the variances of the three distributions, and their sensitivity to parameter change. The Normal's variance is directly parametrized by the action network. In contrast, the variance of the beta distribution depends on both shape parameters $\alpha$ and $\beta$:
$$
Var_\beta = \frac{\alpha \beta}{(\alpha + \beta)^2 (\alpha + \beta + 1)}
$$
which means large, simultaneous increases in both $\alpha$ and $\beta$ is required to lower variance, leading to slow convergence.
In contrast, the Cauchy distribution famously does not have a theoretical mean, variance or kurtosis, due to laws of integration.

This points to why the Cauchy distribution works better, all things held equal. It is parameterized as directly as the normal distribution, but is also has "heavy-tailed" and "peakier" than the Gaussian. This slower decay in the tails means lower variance sensitivity, hence avoiding convergence to local optima. At the same time, the Cauchy distribution also has a much higher peak than the Gaussian, which means there is less cost in accuracy when sampling.

In regards to our study's objectives: the results from Sections~\ref{sec:results-acceptance-strategy} and Section~\ref{sec:results-offer-strategy} show slow convergence when variance is high and sub-optimal convergence when variance is low due to lack of action exploration as the primary limitation. This suggests the learning rate can be adjusted through marginal utilities, the distribution's kurtosis (peaky-ness and tail-behavior), and the variance sensitivity for faster convergence and efficient outcomes. A more aggressive learning rate can curtail distributions with lower variance sensitivity. Furthermore, variation in concession factor and discount rate yields different strategies from Offer Net, thus demonstrating adaptivity.

\section{Self-Play: The Emergence of Fairness}
So far, we have addressed sub-problems related to training barriers and demonstrated exploitative capabilities against time-based agents.
While play against time-based agents provides clear benchmarks due to monotonic time-based concession, play against behavior-based agents is required to evaluate behavioral traits such as fairness. 
In this section, we first present a game-theoretic framework of our games, then the results for single- and multi-issue self-play, then against two variants of tit-for-tat agents.

\subsection{Game-theoretic Framework} \label{sec:game-theory}
We introduce a few game-theoretic concepts required for in-depth behavioral analysis. An \textit{extensive game} consists of a set of players $N$, a set of sequences $H$ that denote possible game trajectories. A \textit{game tree} describes this trajectory of states, round-by-round.  
A \textit{Nash Equilibrium} (NE) denotes an outcome where no player wants to willingly deviate. In extensive games, a strategy profile $s*$ is an NE if
$$O(s^*{-1},s^*_i) \geq O(s^*{-1},s_i) \qquad \forall s_i \in S_i$$
$S_i$ denotes the strategy set of player $i$. Let $(<a_1,a_2,...>,<b_1,b_2,...>)$ denote the strategy profiles, where each bracket contains a player's sequence of moves~\cite{osborne1994course}. 

\begin{figure}[!htbp] 
\centering
\caption{Centipede Game\label{fig:centipede-game_tree}}
\begin{tikzpicture}[font=\footnotesize,scale=1]
\tikzstyle{solid node}=[circle,draw,inner sep=1.2,fill=black];
\tikzstyle{hollow node}=[circle,draw,inner sep=1.2];
\node(0)[hollow node]{}
child[grow=down]{node[solid node]{}
	edge from parent node[left]{$D$}}
child[grow=right]{node(1)[solid node]{}
child[grow=down]{node[solid node]{}
	edge from parent node[left]{$D$}}
child[grow=right]{node(2)[solid node]{}
child[grow=down]{node[solid node]{}
	edge from parent node[left]{$D$}}
child[grow=right]{node(3)[solid node]{}
child[grow=down]{node[solid node]{}
	edge from parent node[left]{$D$}}
child[grow=right]{node(4)[solid node]{}
child[grow=down]{node[solid node]{}
	edge from parent node[left]{$D$}}
child[grow=right]{node(5)[solid node]{}
child[grow=down]{node[solid node]{}
	edge from parent node[left]{$D$}}
child[grow=right]{node(6)[solid node]{}
edge from parent node[above]{$C$}}
edge from parent node[above]{$C$}}
edge from parent node[above]{$C$}}
edge from parent node[above]{$C$}}
edge from parent node[above]{$C$}}
edge from parent node[above]{$C$}};
\foreach \x in {0,2,4}
\node[above]at(\x){1};
\foreach \x in {1,3,5}
\node[above]at(\x){2};
\node[below]at(0-1){($0.9,0.1$)};
\node[below]at(1-1){($0.2,1.8$)};
\node[below]at(2-1){($2.7,0.3$)};
\node[below]at(3-1){$(0.4,3.6)$};
\node[below]at(4-1){$(4.5,0.5)$};
\node[below]at(5-1){$(0.6,5.4)$};
\node[right]at(6){$(3.5,3.5)$};
\end{tikzpicture}
\caption{Game tree of the centipede game. Every round, the "pie" grows by $1$, ending with $7$. 
Players can split it evenly at the end (cooperate every turn), or defect. 
By backwards induction, the SPNE is $(<D,D,D>,<D,D,D>)$, with reward$=(0.9,0.1)$.}
\end{figure}

In extensive games, the concept of \textit{sub-games} describes part of the game tree which function as a game itself~\cite{osborne1994course}. Fig.~\ref{fig:centipede-game_tree} shows the game tree of the \textit{centipede game}, a canonical game in game theory, and Fig.~\ref{fig:bargaining-game_tree} shows the game tree of a bargaining game.  
In this instance of the centipede, the total size of the pie increases by $1$ at every time step. The players can choose to wait or defect. Consider the rightmost node in Fig.~\ref{fig:centipede-game_tree} labeled $2$, denoting P2's decision to cooperate or defect. Since the pay-off of defecting yields a reward of $5.4$ over $3.5$ from cooperating, P2 will defect if they are rational. The sub-tree stemming from $2$ can then be reduced to $(0.6,5.4)$.

Once P1 realizes P2 will defect, P1 will also defect as this yields a higher reward. This process continues until P1 defects in round 1. The process of iteratively reducing up the tree is known as \textit{backwards-induction}. The result at the end is a \textit{sub-game perfect Nash Equilibrium} (SPNE), a type of NE that is also the equilibria of sub-games. The SPNE in this game is found to be $<D,D,D>,<D,D,D>$. Ironically, if both players waited until the end, they would receive higher rewards. Hence, for the centipede game, we expect to see cooperative agents wait until the end, while rational agents defect at the beginning.

\begin{figure}[!htbp] 
\centering
\caption{Bargaining Game\label{fig:bargaining-game_tree}}
\begin{tikzpicture}[font=\footnotesize,scale=1]
\tikzstyle{solid node}=[circle,draw,inner sep=1.2,fill=black];
\tikzstyle{hollow node}=[circle,draw,inner sep=1.2];
\node(0)[hollow node]{}
child[grow=down]{node[solid node]{}
	edge from parent node[left]{$R,O$}}
child[grow=right]{node(1)[solid node]{}
child[grow=down]{node[solid node]{}
	edge from parent node[left]{$D$}}
child[grow=right]{node(2)[solid node]{}
child[grow=down]{node[solid node]{}
	edge from parent node[left]{$D$}}
child[grow=right]{node(3)[solid node]{}
child[grow=down]{node[solid node]{}
	edge from parent node[left]{$D$}}
child[grow=right]{node(4)[solid node]{}
child[grow=down]{node[solid node]{}
	edge from parent node[left]{$D$}}
child[grow=right]{node(5)[solid node]{}
child[grow=down]{node[solid node]{}
	edge from parent node[left]{$D$}}
child[grow=right]{node(6)[solid node]{}
edge from parent node[above]{$C$}}
edge from parent node[above]{$C$}}
edge from parent node[above]{$C$}}
edge from parent node[above]{$C$}}
edge from parent node[above]{$C$}}
edge from parent node[above]{$C$}};
\foreach \x in {0,2,4}
\node[above]at(\x){1};
\foreach \x in {1,3,5}
\node[above]at(\x){2};
\node[below]at(0-1){($x_1,y_1$)};
\node[below]at(1-1){($x_2,y_2$)};
\node[below]at(2-1){($x_3,y_3$)};
\node[below]at(3-1){$(x_4,y_4)$};
\node[below]at(4-1){$(x_5,y_5)$};
\node[below]at(5-1){$(x_6,y_6)$};
\node[right]at(6){$(0,0)$};
\end{tikzpicture}
\caption{Game tree of a bargaining game. With an initial size of $1$, the size diminishes by $0.9$ each time step, ending with $0$ after the sixth and last round, when the conflict deal is enacted. The sum of the rewards are subject to $x_i + y_i = 0.9^{t-1}$.}
\end{figure}

\subsection{Univariate self-play results} \label{sec:uni-self-play}
Before training with self-play in the multi-issue domain, we consider a simplified version. Instead of giving offers in $[0,1]^3$, consider the case where bidding actions are constrained to a binary decision---either offering $(0.5,0.5)$ or $(0.9,0.1)$.
Thus, agents can either offer a low amount to their opponent (rational behavior), or a fair amount. These four choices can be summarized as:
\begin{enumerate}
\itemsep-0.5em 
    \item [$(L,L)$: ] Offer low, reject nothing. This is typically the SPNE, thus \textbf{rational (G1)}.
    \item [$(H,L)$: ] Offer high, reject nothing. This is \textbf{altruistic (G2)}.
    \item [$(H,H)$: ] Offer high, accept high. This agent is \textbf{fair (G3)}.
    \item [$(L,H)$: ] Offer low, accept high. This one is often disregarded, as it is a hardliner \textbf{G4}.
\end{enumerate}

\begin{figure}[!htb]
    \centering
    \includegraphics[width =1.0\linewidth]{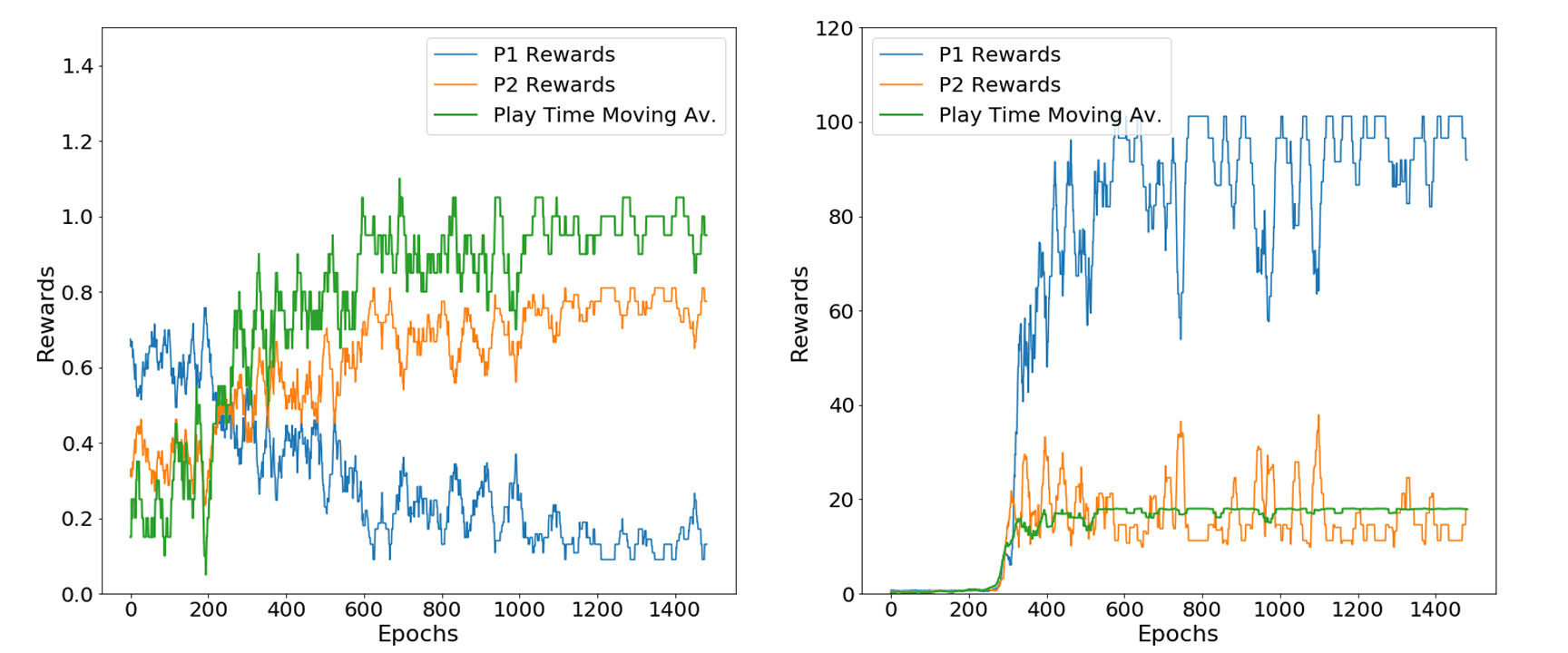}
    \caption{Training results for self-play, for the centipede game and bargaining game. Depicted are P1's results(blue), P2's results(orange) and the total playing time (green).}
    \label{fig:self-play-training}
\end{figure}

 Fig.~\ref{fig:self-play-training} shows the training results for the bargaining and centipede game, with the discount factors set to $0.9$ and $1.3$ respectively. Note, by setting the discount rate to greater than $1$, the bargaining game effectively becomes a more complex version of the centipede game.
For the bargaining game in Fig.~\ref{fig:self-play-training}a), the reward for P2 is initially low, then increases with time approaching 1 round. Conversely, the reward for P1 decreases. We infer that P2 learns to reject P1's offer, and P1 learns to accept. This play is close to rationally optimal. If the agents were perfectly rational, the game would end immediately. However, P2 adopts a strategy that forces play to go on--- we analyze the reason for this further in the multivariate case.

\begin{figure}[!htb]
    \centering
    \includegraphics[width =1.0\linewidth]{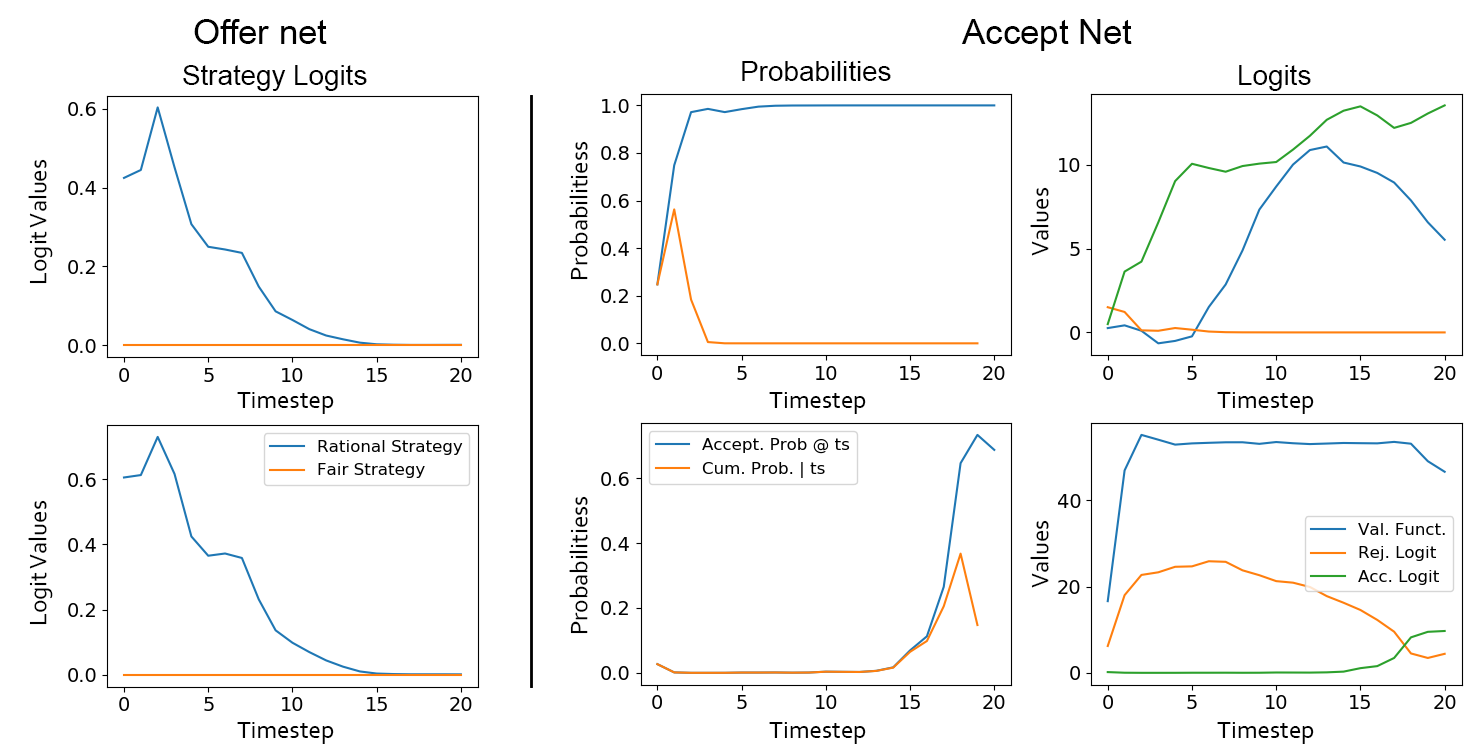}
    \caption{Comparison of offer and acceptance logits for the (a) bargaining and (b) centipede game. In both, Offer Net initially gives rational offers, then over time shifts to fair offers to increase acceptance probability. Accept Net accepts early in the bargaining game, and late for the centipede game, aligned with the optimal stopping times.}
    \label{fig:self-play-logits}
\end{figure}

 In the centipede game in Fig.~\ref{fig:self-play-training}b), the players learn to play close to 20 rounds, maximizing the "interest" accumulated. The total size of the pie is $1.3^{19} = 146$. By the final round, P1 holds a mixed strategy yields 47\% rational offers ($0.9,0.1)$ and 53\% fair offers ($0.5,0.5)$. As the time series shown are the running averages, the big dips show brief spans where the P1 adopts a fair strategy.
These dynamics can be seen more clearly by observing how decision logits evolve during gameplay. Fig.~\ref{fig:self-play-logits} shows the probabilities of giving rational and fair offers at each time step, and the probabilities of accepting an offer (for one game trajectory).

In both cases, agents start-out giving low offers to their opponents. However, as time moves forward, the probability of a fair offer increases, to increase the probability of acceptance. For the bargaining game (a), the stoppage time reaches a maximum close to the beginning. This suggests the network, through gameplay, learns outcomes similar to backward-induction.
Similarly in the centipede game, the network learns to wait, leveraging ``interest" to accept near to the deadline, which also indicates \textit{cooperative behavior}. Together, we conclude the neural agent learns to accept optimally and as time moves forward, shift its behavior from G1 (rational) to G3 (fair).

\subsection{Multivariate Self-play}  \label{sec:multi-self-play}
\begin{figure}[!htb]
    \centering
    \includegraphics[width =0.8\linewidth]{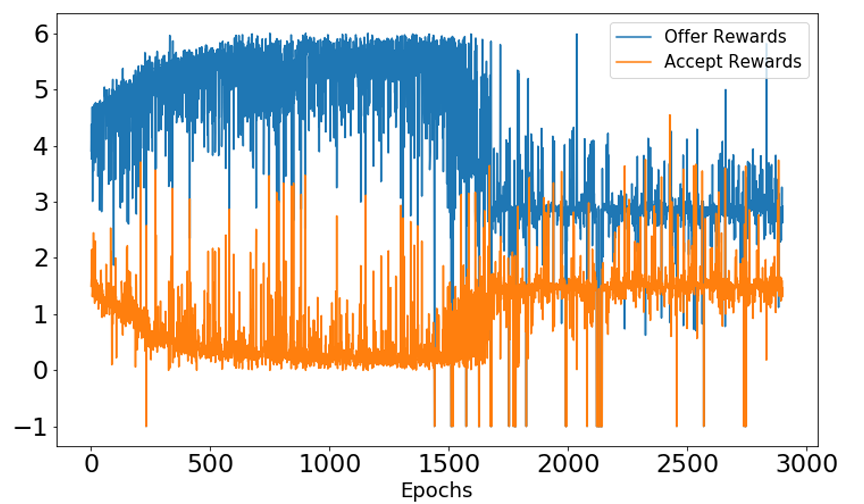}
    \caption{Multivariate self-play for negotiation. Offer Net concedes value after Accept Net adopts a mixed strategy with the conflict deal threat.}
    \label{fig:multi-selfplay-rewards}
\end{figure}

Now, we extend analysis to the continuous, multi-issue case.
Fig.~\ref{fig:multi-selfplay-rewards} shows the training dynamics of the Offer Net and Accept Net, whose final rewards are plotted per epoch, with the discount rate set to $0.95$. Initially, the Offer Net (blue) has higher reward--- as long as some reward is given to Accept Net, the Accept Net will accept it. However, around epoch 1600 the Accept Net learns to invoke the conflict deal. This is demarcated by the large drop in reward for both blue and orange to $-1$. After which, the Offer Net must concede some by offering a fairer amount. 

In sum, by including some probability of the conflict deal, neural agents force a counter-offer that is fairer. This departs from classical game theory, as an example of a \textit{non-credible threat}. A non-credible threat describes actions that perfectly rational agents will not carry out, as it would also leave themselves worse off~\cite{osborne1994course}. 
The adaption of non-credible threats is also observable in the uni-variate centipede game, with periods of dips in P1's reward, following the conflict deal. For discounted bargaining, convergence to low playing time suggests that heavy discounting acts as a similar threat--- if you do not offer a fair deal, I will drag on the negotiation. By keeping non-credible threats part of a mixed strategy, fair outcomes can evolve. 

This is significant because it agrees with results from evolutionary game theory. We previously mentioned in Section~\ref{sec:game-theory}, that reputation produces fairness in the repeated ultimatum game. Nowak et al. showed this through the same, exact simplified mini-game (bids restricted to low and fair offers), then full bidding space using population-based experiments~\cite{nowak2000fairness}. Populations of G1, G2, and G3 played against each other, and ``reproduced" based on their utility. After many rounds, results showed that the rational agent (G1) dominated. However, if the prior acceptance history was available, which showed agents rejecting below a certain threshold, then a population of fair agents (G3) who offered high and accepted high would emerge.

In other words, there is a strong similarity between non-credible threats in our neural agent and the rejection of low offers (using reputation) in evolutionary strategies. This is by far the most interesting result, and it's important to note evolutionary methods and RL are often framed as competing choices for agent design~\cite{salimans2017evolution}. Similar results produced in these two fields may drive future research directions.


\subsection{Against Tit-for-Tat Agents}  \label{sec:results-t4t}
\begin{figure}[!htb]
    \centering
    \includegraphics[width =.95\linewidth]{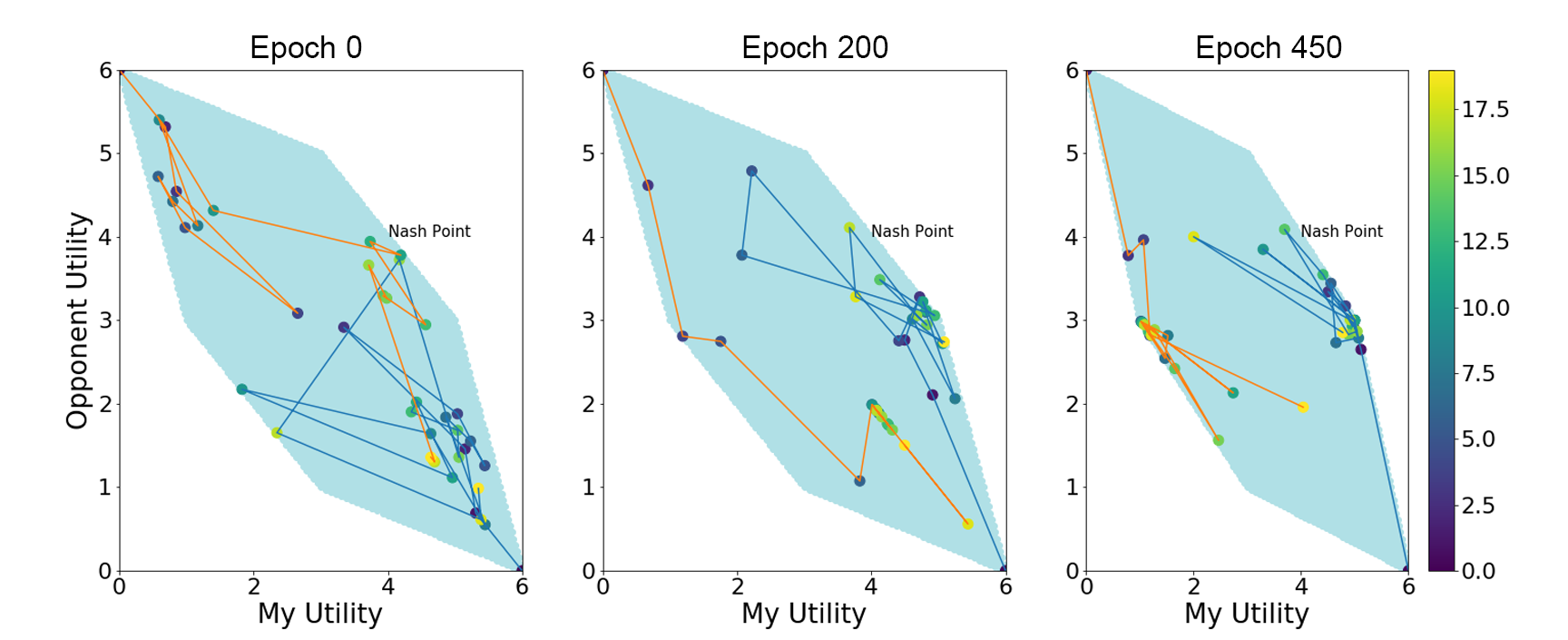}
    \caption{Evolution of play against relative tit-for-tat agents.    
}
    \label{fig:t4t-vanilla}
\end{figure}
Finally, we present results against relative TFT and the Bayesian TFT agent. 
Since this investigation studies whether the acceptance and bidding strategy can adapt and induce promising counter-bids, 
each game is designed as follows: the TFT agent makes a bid, then the neural agent makes an acceptance decision and counter-bids. Thus, the game can only end on the neural agent's acceptance.

Against the relative TFT agent with no discounting (Fig.~\ref{fig:t4t-vanilla}), the neural agent converges to the $(0,1,1)$ vertex. Yellow represents the ending epoch and we observe the neural agent's utility is greater than 3. Notably, the TFT agent makes offers opposite of the Pareto Frontier. This arises as the relative TFT agent measures concession with respect to its own utility. As we've analyzed in the time-based opponents, DRL agents are prone to adjusting variables that yield the greatest rewards and lowest losses. To the TFT agent, this is reversed, prompting it to concede the issue it values most and propose away from the Pareto frontier. This demonstrates \textit{adaptivity}.

\begin{figure}[!htb]
    \centering
    \includegraphics[width =1.0\linewidth]{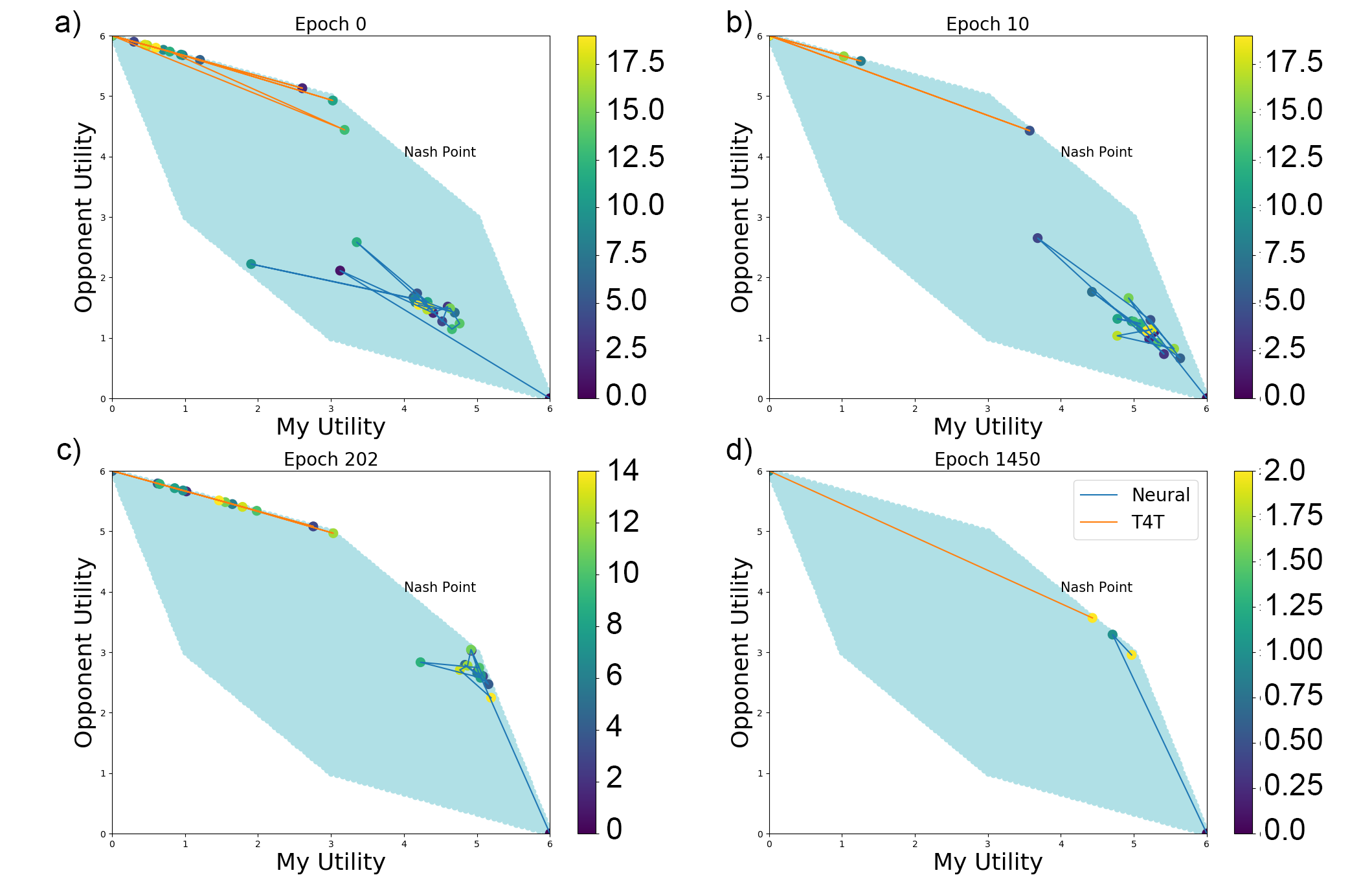}
    \caption{Evolution of play against Bayesian tit-for-tat agents. The agent learns to accept early and around the Nash point.}
    \label{fig:Bayesian-t4t}
\end{figure}

In contrast, the neural agent \textit{cooperates} with the Bayesian TFT agent. In Fig.~\ref{fig:Bayesian-t4t}a (epoch 0) the neural agent performs randomly and suboptimally. Note, the color bar represents time step. However, the bid direction drifts towards the $(0,1,1)$ vertex (Figs~\ref{fig:Bayesian-t4t}(b) and (c)). By epoch 1450, the neural agent learns to induce results near the Nash point, in only four moves due to the discount rate.

With results from time-based agents and self-play, our analysis shows that when concession is necessary, the bidding strategy gravitates toward to $(0,1,1)$. This ensures near Pareto Optimal payoff if its offer is accepted. Exploitation occurs against time-based and relative TFT agents, while fairer outcomes arise with more complex agents.
\chapter{Conclusion}
\section{Summary of Discussion}
Bilateral negotiation presents a unique domain that combines discrete and continuous control problems. Furthermore, the deadline produces a utility function analogous to cliff-walking. This paper is a fundamental evaluation of actor-critic models for negotiation, measuring its ability to exploit, adapt, and cooperate.

The neural agent shows clear \textbf{exploitative behavior} against time-based agents. For acceptance, the neural agent demonstrates precise logit switching behavior, in transitions between rejecting and accepting offers. 
The acceptance strategy resembles a conservative agent, accepting a little before the optimal time. 
For the bidding strategy, we evaluated the use of Normal, Cauchy, and beta distributions for continuous control. The Cauchy has the highest reward, but the Normal is more consistent. The neural agent learns ways to evaluate the opponent, such as maintaining high mean, high initial variance to ensure enough rejections, before lowering the variance to more deterministic outcomes. This also demonstrates adaptability to concession and discounting. 

Time-based experiments reveal the \textbf{barriers} to optimal convergence. We discover the error in stoppage time can be explained by the change in marginal utility (second derivative) and cliff-walking: the agent waits for higher rewards, then is punished aggressively due to enacting the conflict deal. The primary factors that influence the bidding optimality is trade-offs in variance (i.e.the beta distribution suffers from slow convergence due to low variance sensitivity). High variance is required to seek out optimal strategies, but low variance helps avoid the conflict deal. The peakiness of the Cauchy and its heavy tails makes it a suitable candidate.

The neural agent was also shown to be \textbf{cooperative} and \textbf{adaptive}. When playing against time-based agents with preference-based concessions, offers are accepted along the Pareto Frontier and produce the highest expected reward. Self-play in the centipede game shows agents are willing to accrue interest, which demonstrates cooperation over rationality. Against simple Bayesian TFT agents, the neural agent learns to quickly arrive at the Nash Solution, resulting in win-win cooperation. Since all results arise from a single neural architecture, the neural agent shows significant adaptability.
Most importantly, the neural agent forces fairer results by either 1) utilizing the conflict deal or 2) levying discounting to force fairer offers. There is a strong similarity between non-credible threats in our neural agent and the rejection of low offers (using reputation) in evolutionary strategies. It's important to note evolutionary methods and RL are often framed as competing choices for agent design~\cite{salimans2017evolution}. Beyond theoretical interest in diverging from classical game theory, these results may guide the design of fairer negotiations, with EGT from a population perspective and DRL from the individual agent's perspective.

Before discussing future work, I'll note what didn't work. Initially, the use of LSTMs seemed promising due to its success in natural language negotiation generation. However, there the action domain is discrete and limited. 

\subsection{Evaluation and Future Work}
One weakness of this study is it studies a specific preference ordering. For the \textbf{scenario}, promising avenues include variations in the utility functions, as there are six combinations of preference orderings for three issues. More importantly is the inclusion of more complicated behavior-based agents. One barrier to this is, unlike the iterated prisoner dilemma that has hundreds of established strategies, we lack a repository that collects these strategies, such as the Axelrod library for the IPD~\cite{axelrodproject}. 

However, this is quickly changing. An annual negotiation competition that began in 2010~\cite{baarslag2012first} collects strong bots into the Genius Environment~\cite{lin2014genius}, maintained by Tim Baarslag. I anticipate running the neural agent against these bots, to understand how DRL performs in a tournament setting and against more complicated strategies. 

Another weakness of this study is experimentation with design choices (\textbf{learning methodology}), although this was not possible given the focus of this dissertation was behavioral analysis. A separate study with a deep learning focus could scope-out the impact of neural architecture (the type of non-linearity and number of layers) and hyper-parameters (learning rate, reward discounting and the use of schedulers). Additionally, increasing the complexity of the algorithm may improve performance, such as increasing the input space to include $n$-prior moves against trajectory-based opponents, or the use of Monte-Carlo Tree Search and rollout~\cite{silver2016mastering}.

\bibliographystyle{plain}
\bibliography{mybibfile}
\appendix
\chapter{Appendix}
\section{Policy-Gradient Theorem} \label{sec:policy-gradient-theorem}
The policy gradient theorem states the change in scalar is proportional to change in policy weights. More specifically, this is given as:
$$
\nabla J(\theta) \propto \Sigma_s \mu(s) \Sigma q_\pi (s,a) \nabla \pi (a | s,\theta)
$$Here, $\mu(s)$ denotes the on-policy distribution of a state under policy $\pi$. We can think of this as the frequency a state has occurred.  $q_\pi (s,a)$ is the value of the state-action pair and $\nabla \pi (a | s,\theta)$ is the change in the policy distribution. What this says is a positive change in $J(\theta)$ can be produced a proportional shift in the policy. A full proof can be found in Chapter 13 of ~\cite{sutton2018reinforcement}.

\section{Point to Line Calculation} \label{sec:point-2-line}
The general form of the closest distance from a point to line is
\begin{equation}
    d(ax+by+c=0, (x_0, y_0)) = \frac{|ax_0+by_0+c|}{\sqrt{a^2+b^2}}. 
\end{equation}

The Pareto Frontier s given by Eq.~\ref{eq:pareto-frontier}. Hence, the distance of a point $P = (x_p, y_p)$ to the Pareto Frontier is:
$$
\min  \Bigg\{   \frac{|x_p + \frac{1}{3} y_p - 6|}{\sqrt{\frac{10}{9}}}  ,  
							\frac{|x_p + y_p - 8|}{\sqrt{2}}   ,
                            \frac{|x_p + 3 y_p - 18|}{\sqrt{10}}  
\Bigg\}
$$

\section{Actor-Critic Playout Implementation}
\begin{algorithm}[H] \label{algo:training}
\SetAlgoLined
 \For{e in epochs} {
      \While{Not accepted and $t < $deadline}{
        P1 offers\;
        \eIf{P2 Accepts}{
         Collect States for both players \;
         Collect Acceptance Actions and Rewards for both players\;
         }{
         P2 Offers\;
         Swap Places (P1 receives then counter offers)\;
         $t += 1$\;
        }
      }
      Calculate Critic and Actor Loss for both Accept Net and Offer Net\;
      Backprop on both networks\;
  }
 \caption{Implementation of negotiation playout and training pipeline}
\end{algorithm}

\section{Change in Univariate Mean Estimation}
The y-axis shows the decision utility at a given time, which is inversely related to the concession factor. The higher the decision utility, the more Boulware the agent is. The concession value can be converted with $c = \frac{\ln \frac{t}{T}}{\ln U_d}$.

\begin{figure}[!htb]
    \centering
    \includegraphics[width = 0.6\linewidth]{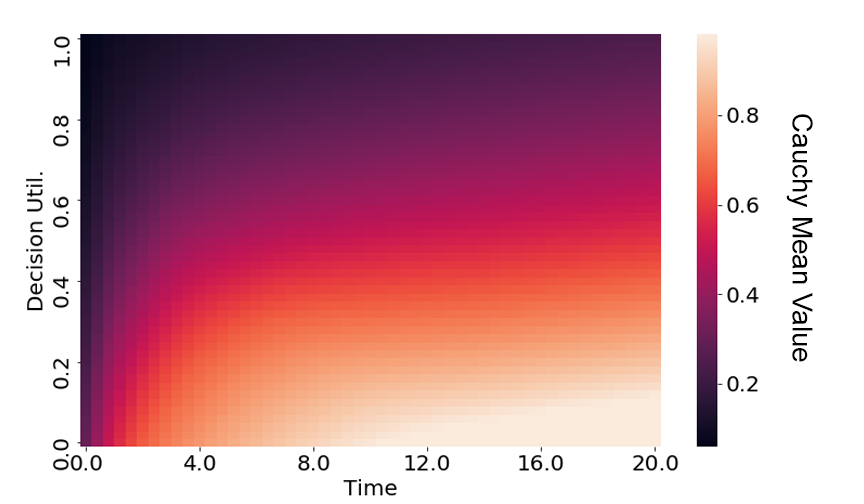}
    \caption{Cauchy means based on opponent decision utility and time. 
    The decision utility serves as a proxy for concession, as the higher the decision utility, the more Boulware the opponent. }
    \label{fig:cauchy-heat}
\end{figure}

\end{document}